\documentclass[%
 reprint,
 amsmath,amssymb,
 aps,
pre
]{revtex4-2}

\usepackage{graphicx}
\usepackage{dcolumn}
\usepackage{bm}

\graphicspath{ {./images/} }
\usepackage{wrapfig}
\usepackage{enumitem}
\usepackage{amsmath,stackengine}
\stackMath
\usepackage{amssymb}
\usepackage{marginnote}
\usepackage{comment}
\usepackage{hyperref}
\usepackage{url}
\usepackage{esint}
\usepackage{multirow}
\usepackage{caption}
\usepackage{subcaption}
\usepackage{tikz}
\usetikzlibrary{arrows.meta}

\newcommand{\e}[1]{^{[#1]}}

\usepackage{mathtools}
\usepackage{listings}
\usepackage{xcolor}
\usepackage{siunitx}
\usepackage{pgfplots}
\pgfplotsset{compat=1.9, , every axis/.append style={font=\scriptsize}}
\usepackage{xspace}

\newcommand{\mb}{\mathbf}

\begin{document}

\title{Asymmetric coupling of nonchaotic Rulkov neurons: \\ Fractal attractors, quasimultistability, and final state sensitivity}

\author{Brandon B. Le}
\email{Contact author: sxh3qf@virginia.edu}
\affiliation{
Department of Physics, University of Virginia, \\
Charlottesville, Virginia 22904-4714, USA
}

\date{\today}

\begin{abstract}
Although neuron models have been well studied for their rich dynamics and biological properties, limited research has been done on the complex geometries that emerge from the basins of attraction and basin boundaries of multistable neuron systems. In this paper, we investigate the geometrical properties of the strange attractors, four-dimensional basins, and fractal basin boundaries of an asymmetrically electrically coupled system of two identical nonchaotic Rulkov neurons. We discover a quasimultistability in the system emerging from the existence of a chaotic spiking-bursting pseudo-attractor, and we classify and quantify the system's basins of attraction, which are found to have complex fractal geometries. Using the method of uncertainty exponents, we also find that the system exhibits extreme final state sensitivity, which results in a dynamical uncertainty that could have important applications in neurobiology.
\end{abstract}

\maketitle


\section{Introduction}
\label{sec:introduction}

Rulkov maps \cite{rulkov, rulkov2} are well established to be simple discrete-time systems that accurately model the behavior of biological neurons. Although many continuous-time neuron models have been developed \cite{hh, chay, buchholtz, izhikevich-model, fh}, the discrete-time Rulkov models have been shown to have a computational advantage over these systems of nonlinear differential equations in modeling complex neuronal systems \cite{izh-time}. Like many other established neuron models \cite{izhikevich-model, hindmarsh, rinzel, izh-map, courbage, omelchenko}, Rulkov's models, often called the nonchaotic and chaotic Rulkov maps \cite{ibarz}, are slow-fast dynamical systems capable of modeling fast bursts of nonchaotic or chaotic spikes on top of slow oscillations. Although the chaotic Rulkov map has been well studied \cite{vries, luo, min, bao, dePontes}, this study involves the nonchaotic Rulkov map, which has received less attention but also exhibits rich dynamical and geometrical properties. The most direct application of the nonchaotic Rulkov map is in modeling neuronal dynamics \cite{rulkov}, but studies have also been conducted using the nonchaotic Rulkov map for stability analysis \cite{wang2}, symbolic analysis \cite{budzinski}, control of chaos \cite{lopez}, machine learning \cite{ge}, information patterns \cite{njitacke}, and digital watermarking \cite{ding}. In this paper, we explore an asymmetric electrical coupling of two identical nonchaotic Rulkov neurons and the emergence of a quasimultistability, a strange spiking-bursting pseudo-attractor, and extreme final state sensitivity in the asymmetrically coupled Rulkov neuron system.

When dynamical systems with attractors are analyzed, it is standard practice to detail their bifurcations, equilibria and eigenvalues, attractor dimensions, maximal Lyapunov exponents, and routes to chaos. However, less attention is often paid to a system's basins of attraction, which are essential to determining multistability properties and the system's use in practical applications \cite{sprott}. Additionally, the fractalization of the boundaries between basins of attraction can result in final state sensitivity, where a small uncertainty in the initial state of a system results in a large uncertainty in the system's eventual final state \cite{grebogi-final-state}. Even more extreme versions of final state sensitivity include Wada basins \cite{daza, kennedy, nusse} and riddled basins \cite{alexander, ott-riddled}. 

Despite these interesting and relevant geometrical properties, limited research has been done in the basins of attraction and basin boundaries of discrete-time neuronal systems. Some studies have been done with continuous-time neuron models, including multistability in a pair of neurons with time delays \cite{shayer}, heterogeneous multistability and riddled basins in an adaptive synapse-based neuron model \cite{bao-neuron}, and multistability, fractal basin boundaries, and final state sensitivity in a network of theta neurons with time-varying excitability \cite{so}. A few studies have also examined these properties in chaotic Rulkov neuron systems, such as multistability in a system of chaotic Rulkov neurons coupled via a chemical synapse \cite{bashkirtseva-ms}, extreme multistability in a system of two chaotic Rulkov neurons with memristive electromagnetic induction \cite{xu}, and basin stability in a system of two chaotic Rulkov neurons with a chemical synaptic and inner linking coupling \cite{rakshit}. However, as suggested by the name, the chaotic Rulkov map is designed to model the dynamics of chaotic spiking-bursting neurons \cite{rulkov2}, so it is not ideal for the purposes of examining basins of attraction and final-state sensitivity properties that arise from multistability. Instead, we are interested in a neuronal map that lends itself to coexisting qualitatively different behaviors, namely nonchaotic regular spiking and chaotic spiking-bursting neuronal dynamics, which the nonchaotic Rulkov map is suited for. However, the downside of using this map is that, while the fast variable of the chaotic Rulkov map is described by a simple iteration function, the nonchaotic Rulkov map's fast variable is described by a piecewise function [see Eq. \eqref{eq:rulkov_1_fast_equation}], which makes analysis significantly more complex. For this reason, to the best of our knowledge, no prior studies have examined the basins of attraction of a nonchaotic Rulkov neuron system or the basin boundary and final state sensitivity properties of any discrete-time neuronal map.

The motivation of our study is therefore to analyze the complex geometrical properties that emerge from the basins of attraction and basin boundaries of a system of two asymmetrically coupled nonchaotic Rulkov neurons. In this system, we find a chaotic spiking-bursting pseudo-attractor and analyze its fractal geometry. Emerging from this chaotic pseudo-attractor and a separate nonchaotic spiking attractor is a kind of quasimultistability and two basins of attraction. We classify and quantify these basins of attraction in multiple subsets of state space using the method of Sprott and Xiong \cite{sprott}, which has been a popular tool to analyze basins of attraction in recent years \cite{gotthans, li-class, sprott-class, nazarimehr, nazarimehr2, nawachioma, marwan}. Then, we use uncertainty exponents \cite{grebogi-final-state, mcdonald} to analyze the system's extreme final state sensitivity and the fractalization of different subsets of the system's basin boundaries using a Monte Carlo algorithm. This research provides insights into how multistability and fractal geometry can emerge from the basins of discrete-time neuronal systems, as well as how the slow-fast behavior of the nonchaotic Rulkov maps leads to complex and unexpected geometries in state space.

This paper is organized as follows. In Sec. \ref{sec:model}, we detail our asymmetrically coupled nonchaotic Rulkov neuron model and the dynamics and apparent multistability that emerge from it. In Sec. \ref{sec:attractor}, we analyze the geometry of the chaotic spiking-bursting pseudo-attractor by computing its box-counting and Lyapunov dimensions. In Sec. \ref{sec:classification}, we overview the basin classification method and classify the intersection of the basins with a two-dimensional slice of state space (Sec. \ref{sec:2d-slice}) and all of four-dimensional state space (Sec. \ref{sec:4d-basins}). In Sec. \ref{sec:boundary}, we investigate different subsets of the system's basin boundaries using uncertainty exponents and explore the existence of extreme final state sensitivity in the system. Finally, we summarize our results and give some future directions in Sec. \ref{sec:conclusions}.

\section{The model and dynamics}
\label{sec:model}

The nonchaotic Rulkov map is a two-dimensional slow-fast map designed to be a simple model of the neuron. Unlike other neuron models composed of highly nonlinear systems of differential equations, the Rulkov map aims to simplify the complex dynamics of a neuron into a phenomenological model that can be used to better understand collective neuronal dynamics. In this way, a limitation of the Rulkov map is that it isn't a completely accurate neuron model with variables and parameters that correspond with physical values with units. Instead, the Rulkov map mimics the behavior of biological neurons in an attempt to be used to aid in the understanding of more complex collective neuronal phenomena, the mechanisms of which can often be hidden behind the mathematical complexity of more biologically accurate neuron models. The main advantage of the Rulkov map is that, despite its simplicity, it is able to model a diverse variety of neuronal behaviors, including regular spiking, irregular spiking, nonchaotic bursts of spikes, and chaotic spiking-bursting \cite{rulkov}. Therefore, in this study, we use the Rulkov map to discover geometrical complexities and final state sensitivity of neuron systems in a phenomenological sense rather than as a prediction of direct experimental results.

The nonchaotic Rulkov map is defined by the following iteration function \footnote{In the original paper that introduces the Rulkov map \cite{rulkov}, the parameter $\sigma'=\sigma+1$ is used, but we use the slightly modified form from \cite{ibarz}.}:
\begin{equation}
    \begin{pmatrix}
        x_{k+1} \\
        y_{k+1}
    \end{pmatrix}
    =
    \begin{pmatrix}
        f(x_k, y_k; \alpha) \\
        y_k - \mu(x_k - \sigma)
    \end{pmatrix},
    \label{eq:rulkov-map}
\end{equation}
where $f$ is the piecewise function
\begin{equation}
    f(x,y;\alpha) = 
    \begin{cases}
        \alpha/(1-x) + y, & x\leq 0 \\
        \alpha + y, & 0 < x < \alpha + y \\
        -1, & x\geq \alpha + y
    \end{cases}.
    \label{eq:rulkov_1_fast_equation}
\end{equation}
Here, $x$ is the fast variable representing the voltage of the neuron, $y$ is the slow variable, and $\alpha$, $\sigma$, and $\mu$ are parameters. To make $y$ a slow variable, we need $0<\mu\ll1$, so we choose the standard value of $\mu=0.001$. To couple Rulkov neurons together with an electrical current, we use a modified version of Eq. \eqref{eq:rulkov-map} that describes the dynamics of the $i$th coupled neuron $\mb{x}_i$ with the iteration function
\begin{equation}
    \begin{pmatrix}
        x_{i,k+1} \\
        y_{i,k+1}
    \end{pmatrix} = 
    \begin{pmatrix}
        f(x_{i,k}, y_{i,k} + \beta_{i,k};\alpha_i) \\[2px]
        y_{i,k} - \mu x_{i,k} + \mu[\sigma_i + \sigma_{i,k}]
    \end{pmatrix},
    \label{eq:rulkov_coupled_mapping}
\end{equation}
where $\mb{x}_{i,k} = (x_{i,k}, y_{i,k})$ is the state of the system at time step $k$, $\alpha_i$ and $\sigma_i$ are parameters of the neuron $\mb{x}_i$, and $\beta_{i,k}$ and $\sigma_{i,k}$ are coupling parameters defined by
\begin{align}
    \beta_{i,k} &= g_{ji}^e\beta^e(x_{j,k}-x_{i,k}), \label{eq:beta_coup} \\
    \sigma_{i,k} &= g_{ji}^e\sigma^e(x_{j,k}-x_{i,k}), \label{eq:sigma_coup}
\end{align}
where $g^e_{ji}$ is the electrical coupling strength from the adjacent neuron $\mathbf{x}_j$ to $\mathbf{x}_i$ and the coefficients $\beta^e$ and $\sigma^e$ set the balance between the couplings of the neuron's fast and slow variables.

To motivate this choice of coupling where we add values proportional to the voltage difference to the slow variable $y$ and the effective value of $y$ in the fast map, we will briefly discuss the qualitative behavior of the nonchaotic Rulkov map \footnote{For a more in-depth discussion, see Refs. \cite{rulkov} and \cite{bn}.}. From Eq. \eqref{eq:rulkov_1_fast_equation}, we can see that a higher value of the slow variable $y$ effectively raises the height of the fast variable iteration function $f$, which results in a quicker increase in $x$ before being reset by the third piece of $f$. Hence, higher values of $y$ result in shorter periodic orbits or more rapid spikes. It is easy to see from Eq. \eqref{eq:rulkov-map} that the parameter $\sigma$ is the value of $x$ required to keep $y$ constant. If $x<\sigma$, $y$ will slowly increase, and vice versa. Therefore, $\sigma$ is an ``excitation parameter'' for the Rulkov map: a higher value of $\sigma$ will result in $y$ to increase until the spikes are fast enough for the average value of $x$ to equal $\sigma$. From this, it is clear that $\sigma_{i,k}$ acts as a parameter representing gradual changes in dynamics resulting from a voltage difference between neurons, essentially resulting in a slow increase or decrease in the frequency of spikes depending on whether current is flowing in or out of the neuron. On the other hand, $\beta_{i,k}$ represents an immediate change in the dynamics of the neuron as it changes the effective value of $y$ for the slow variable $x$. In other words, if the voltage difference is positive and current is flowing into a neuron, the $\beta_{i,k}$ results in the slow variable $x$ ``thinking'' that the fast variable $y$ is larger, causing an immediate excitation. The effects of both coupling parameters are valuable to our study and have been observed in real biological neurons \cite{rulkov}.

In this paper, we examine a system of two asymmetrically electrically coupled nonchaotic Rulkov neurons, denoted by $\mb{x}_1$ and $\mb{x}_2$. Since these neurons are coupled asymetrically, $g_{21}^e\neq g_{12}^e$, so we denote $g_1^e = g_{21}^e$ and $g_2^e = g_{12}^e$. Setting $\beta^e=\sigma^e=1$ for simplicity, Eqs. \eqref{eq:beta_coup} and \eqref{eq:sigma_coup} read
\begin{align}
    \mathfrak{C}_{1,k} &= \beta_{1,k} = \sigma_{1,k} = g^e_1(x_{2,t}-x_{1,t}), \label{eq:coup_param_1} \\
    \mathfrak{C}_{2,k} &= \beta_{2,k} = \sigma_{2,k} = g^e_2(x_{1,t}-x_{2,t}), \label{eq:coup_param_2}
\end{align}
where $\mathfrak{C}_{i,k}$ is the coupling parameter for both the slow and fast variables of neuron $i$ at time step $k$. 

Since our model is composed of two coupled neurons, each with two variables, the state vector of the model is four-dimensional:
\begin{equation}
    \mathbf{X} = 
    \begin{pmatrix}
        X^{[1]} \\[2px]
        X^{[2]} \\[2px]
        X^{[3]} \\[2px]
        X^{[4]}
    \end{pmatrix}
    = \begin{pmatrix}
        x_{1} \\
        y_{1} \\
        x_{2} \\
        y_{2}
    \end{pmatrix},
\end{equation}
where $X\e{p}$ is the $p$th entry of the state vector $\mb{X}$. This system has the iteration function
\begin{equation}
    \mathbf{X}_{k+1} = \mathbf{F}(\mathbf{X}_k),
\end{equation}
where $\mb{F}(\mb{X})$ can be written explicitly using Eqs. \eqref{eq:rulkov_coupled_mapping}, \eqref{eq:coup_param_1}, and \eqref{eq:coup_param_2}:
\begin{equation}
    \begin{split}
        \mathbf{F}(\mathbf{X}) &= \begin{pmatrix}
            F^{[1]}(x_{1},y_{1},x_{2},y_{2}) \\[4px]
            F^{[2]}(x_{1},y_{1},x_{2},y_{2}) \\[4px]
            F^{[3]}(x_{1},y_{1},x_{2},y_{2}) \\[4px]
            F^{[4]}(x_{1},y_{1},x_{2},y_{2})
        \end{pmatrix} \\
        &= \begin{pmatrix}
            f(x_{1}, y_{1}+g^e_1(x_{2} - x_{1}); \alpha_1) \\[2px]
            y_{1} - \mu x_{1} + \mu[\sigma_1 + g^e_1(x_{2} - x_{1})] \\[2px]
            f(x_{2}, y_{2}+g^e_2(x_{1} - x_{2}); \alpha_2) \\[2px]
            y_{2} - \mu x_{2} + \mu[\sigma_2 + g^e_2(x_{1} - x_{2})]
        \end{pmatrix}.
    \end{split}
    \label{eq:rulkov_1_asym_coup_iter_func}
\end{equation}

The Jacobian matrix $J(\mathbf{X})$ of this system is quite complex due to the piecewise nature of the fast variable iteration function $f$ present in $\mb{F}(\mb{X})$ [see Ref. \cite{bn} for an explicit rendering of $J(\mathbf{X})$], but we can write it in a compact form by partitioning it using the method detailed in Appendix \ref{appx:partition}:
\begin{equation}
    J(\mathbf{X}) = \begin{pmatrix}
        J_{\text{dg},a}(x_{1},\alpha_1,g^e) & J_{\text{odg},b}(g^e) \\
        J_{\text{odg},c}(g^e) & J_{\text{dg},d}(x_{2},\alpha_2,g^e)
    \end{pmatrix}.
    \label{eq:jacobian}
\end{equation}
Using the QR factorization method of Lyapunov exponent calculation \cite{bn, eckmann, sandri, brandon}, we can use this Jacobian matrix to compute the system's Lyapunov spectrum for analysis of the system's dynamics. 

\begin{figure}[t!]
    \centering
    \begin{subfigure}[t]{0.95\columnwidth}
        \centering
        \includegraphics[scale=0.5]{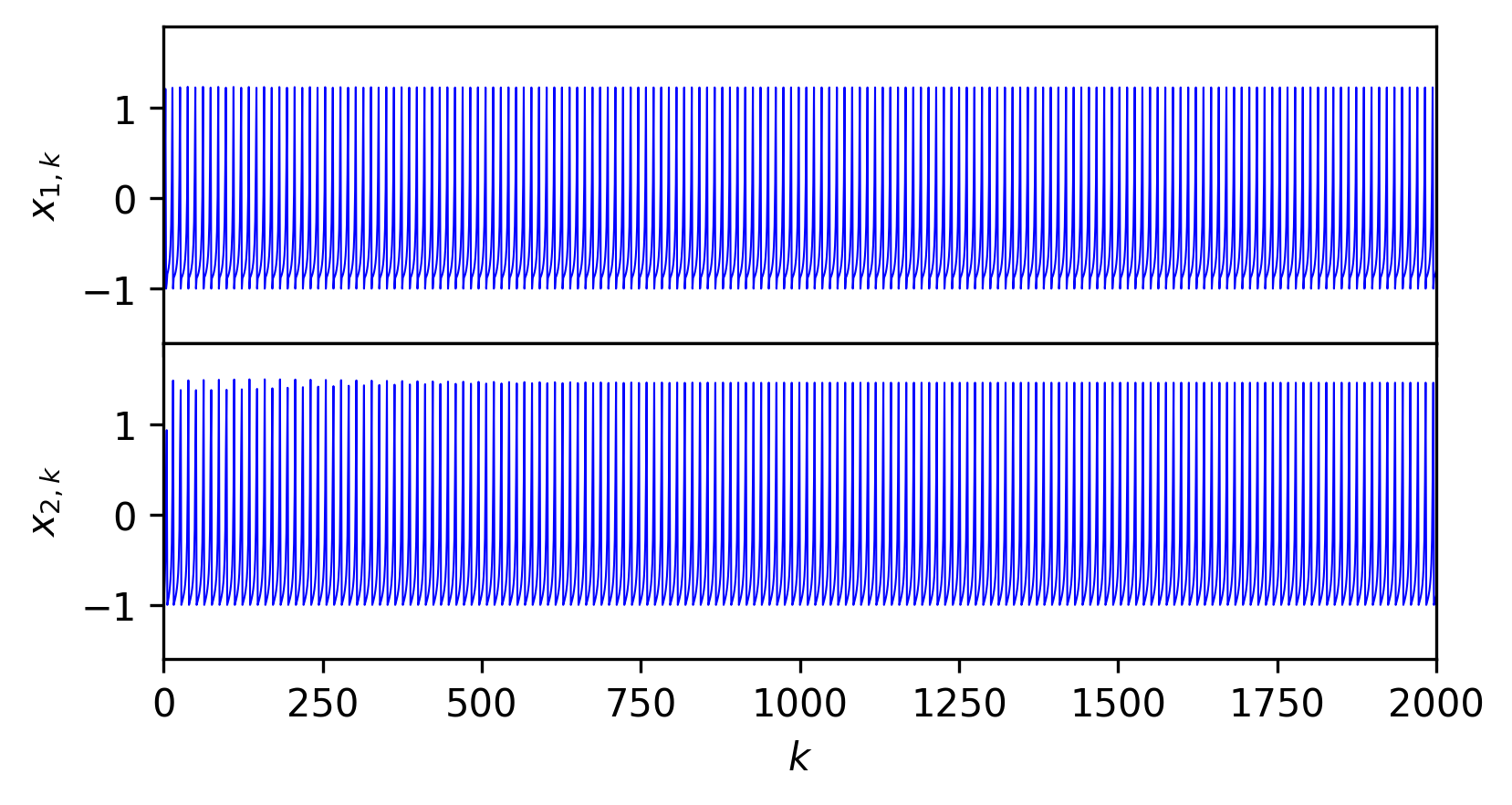}
        \caption{$\mathbf{X}_{0} = (-0.54, -3.25, -1, -3.25)$, $\lambda_1\approx -0.0057$}
        \label{fig:asym_coup_rulkov_1_spiking}
        \vspace{8px}
    \end{subfigure}
    \begin{subfigure}[t]{0.95\columnwidth}
        \centering
        \includegraphics[scale=0.5]{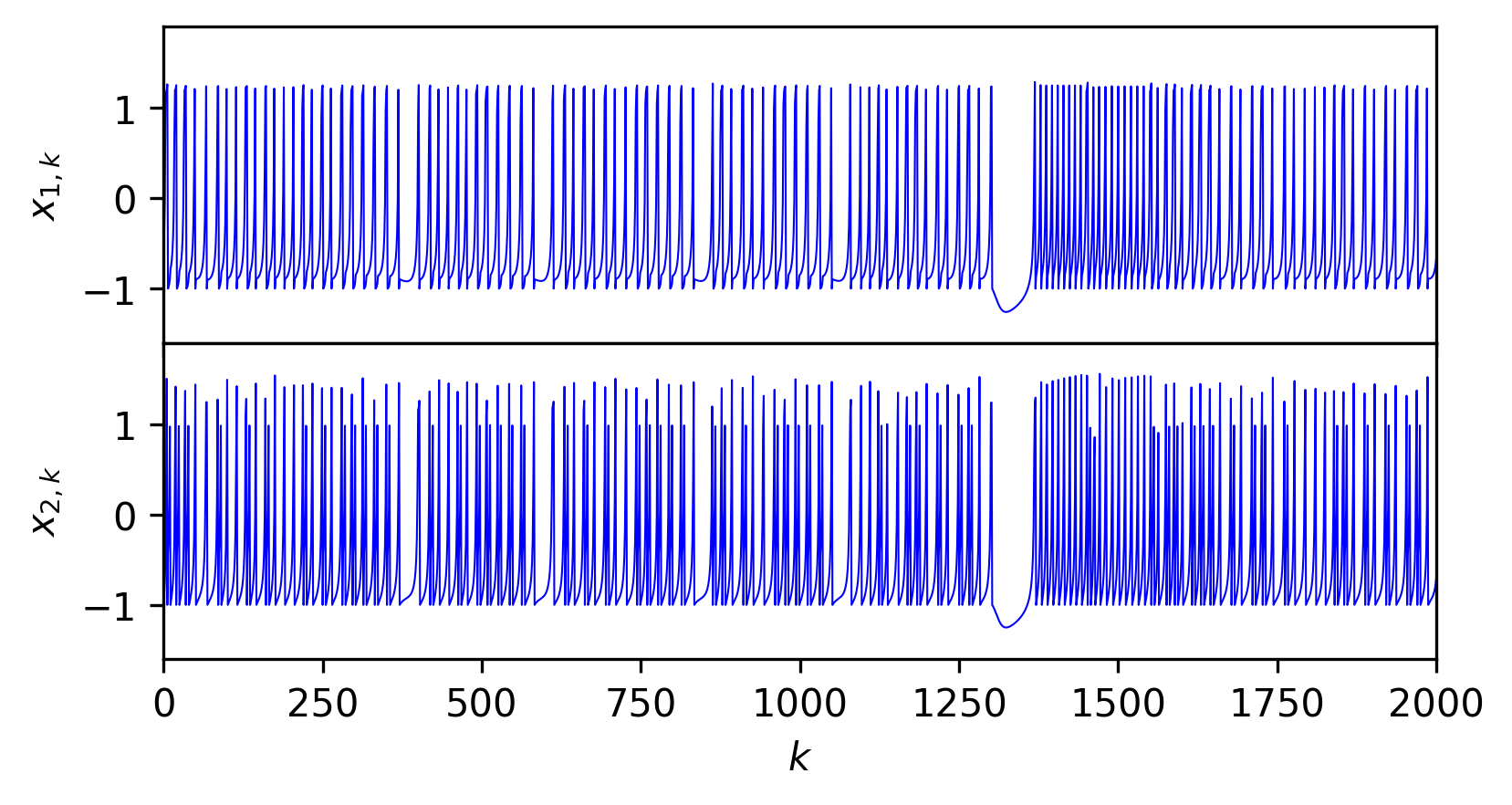}
        \caption{$\mathbf{X}_{0} = (-0.56, -3.25, -1, -3.25)$, $\lambda_1\approx 0.0207$}
        \label{fig:asym_coup_rulkov_1_chaos}
        \vspace{8px}
    \end{subfigure}
    \caption{Graphs of the fast variable orbits $x_{1,k}$ and $x_{2,k}$ of two asymmetrically electrically coupled nonchaotic Rulkov neurons with parameters $\sigma_1=\sigma_2=-0.5$, $\alpha_1=\alpha_2=4.5$, and electrical coupling strengths $g^e_1=0.05$, $g^e_2=0.25$}
    \label{fig:asym_coup_rulkov_1_graphs}
\end{figure}

For our system, we choose the asymmetrically coupled Rulkov neurons to be identical, with parameters $\sigma_1=\sigma_2=-0.5$ and $\alpha_1=\alpha_2=4.5$. This puts each of the individual neurons in the nonchaotic regular spiking domain. The coupling strengths we use are $g^e_1=0.05$ and $g^e_2=0.25$, which results in neuron $\mathbf{x}_2$ ``feeling'' the voltage difference between the two neurons more. To get a snapshot of the dynamics that can emerge from this model, we pick two slightly different initial conditions [$\mb{X}_0 = (-0.54, -3.25, -1, -3.25)$ and $\mb{X}_0 = (-0.56, -3.25, -1, -3.25)$] and plot the fast variable orbits of the two neurons in Fig. \ref{fig:asym_coup_rulkov_1_graphs}. In the first graph [Fig. \ref{fig:asym_coup_rulkov_1_spiking}], the system exhibits nonchaotic, synchronized spiking with a negative maximal Lyapunov exponent $\lambda_1\approx-0.0057$. However, in the second graph [Fig. \ref{fig:asym_coup_rulkov_1_chaos}], after changing the initial value of $x_1$ slightly, chaotic spiking-bursting occurs in both neurons, with a positive maximal Lyapunov exponent of $\lambda_1\approx0.0207$. These two similar initial states immediately being attracted to two drastically different stable orbits, one chaotic and the other nonchaotic, implies the existence of multistability in this system. 

\begin{figure}[t!]
    \centering
    \begin{subfigure}[t]{\columnwidth}
        \centering
        \includegraphics[scale=0.25]{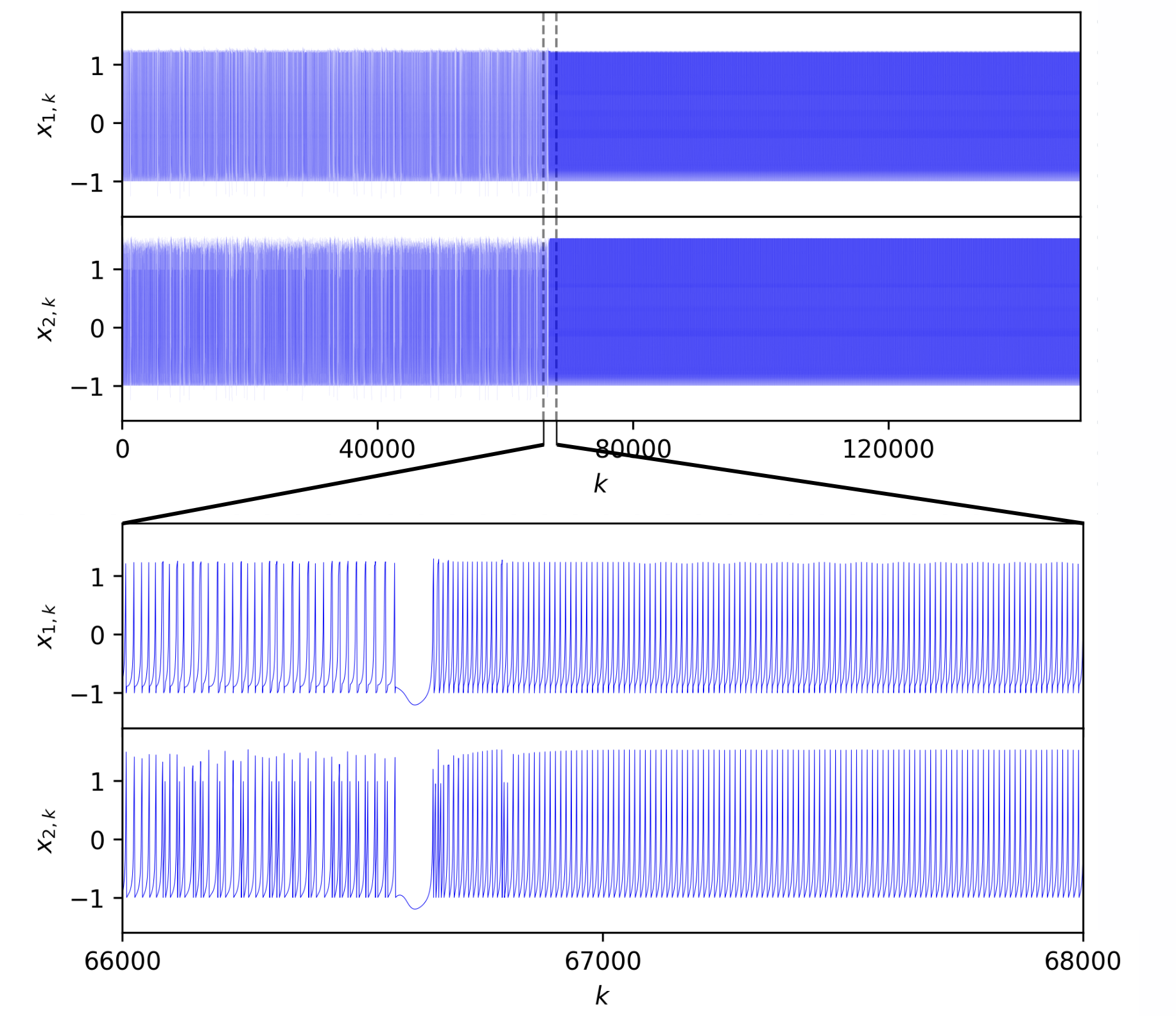}
        \vspace{-2pt}
        \caption{$\mathbf{X}_{0} = (-0.56, -3.25, -1, -3.25)$}
        \label{fig:asym_coup_rulkov_1_long_1}
        \vspace{0.5cm}
    \end{subfigure}
    
    \begin{subfigure}[t]{\columnwidth}
        \centering
        \includegraphics[scale=0.25]{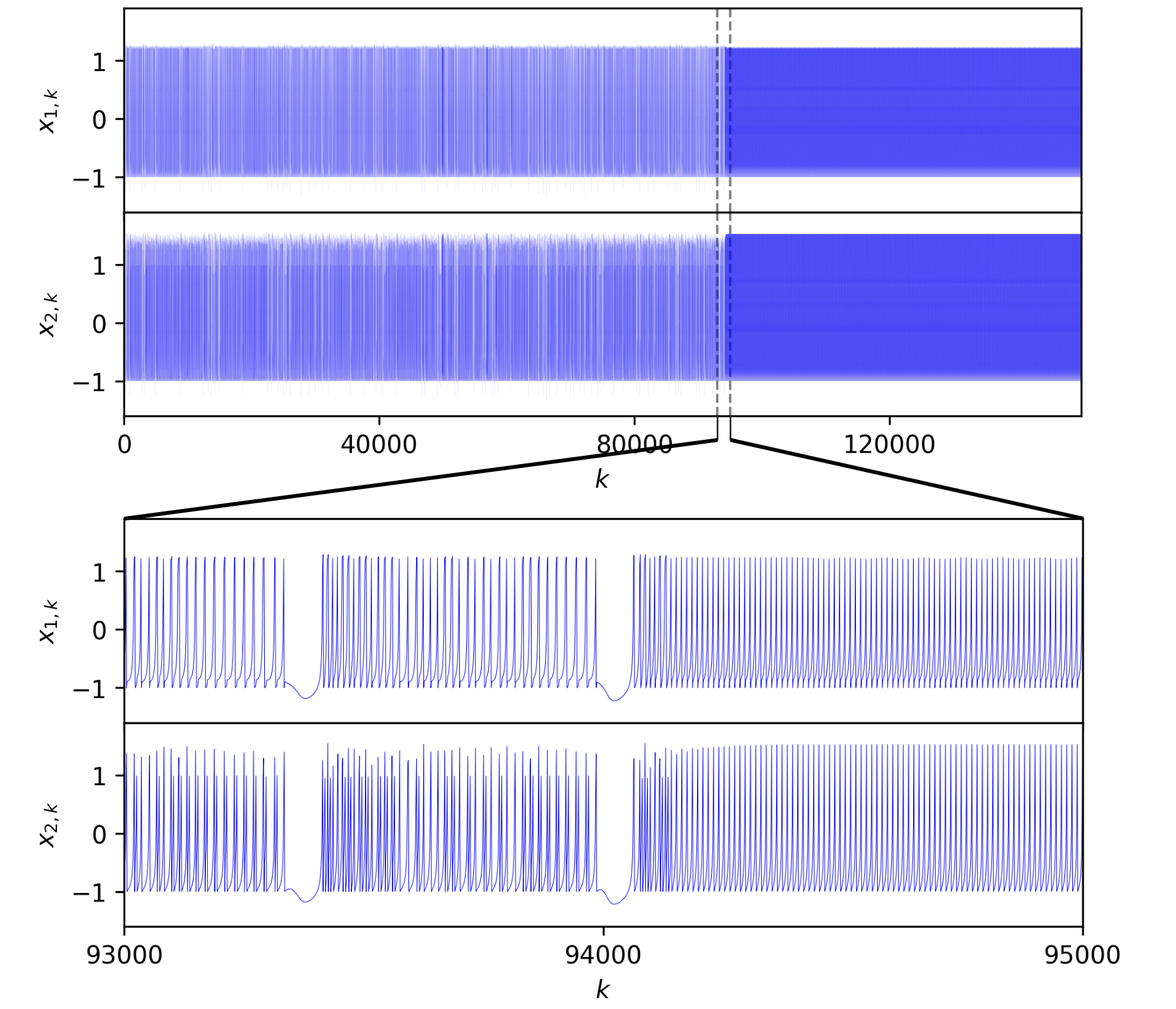}
        \caption{$\mathbf{X}_{0} = (0.35, -3.25, -1.23, -3.25)$}
        \label{fig:asym_coup_rulkov_1_long_2}
    \end{subfigure}
    \caption{Graphs showing the attraction of two chaotic orbits to the system's nonchaotic spiking attractor}
    \label{fig:asym_coup_rulkov_1_long_graphs}
\end{figure}

\begin{figure*}[ht!]
    \centering
    \begin{subfigure}[t]{0.95\columnwidth}
        \centering
        \includegraphics[scale=0.5]{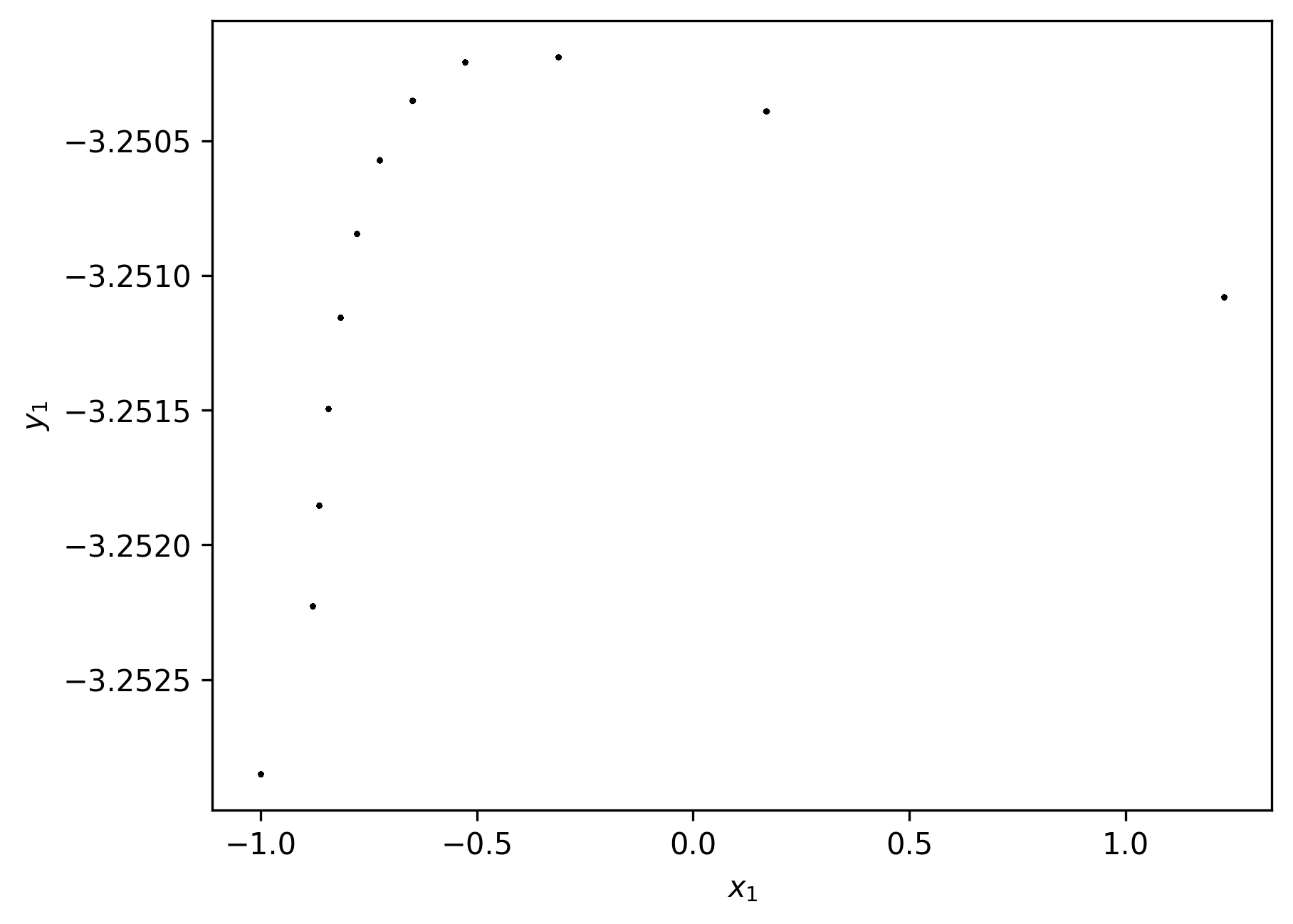}
        \caption{Projection onto the $(x_1,y_1)$ plane}
        \label{fig:nonchaotic_rulkov1_attractor_neuron1}
        \vspace{0.1cm}
    \end{subfigure}
    \begin{subfigure}[t]{0.95\columnwidth}
        \centering
        \includegraphics[scale=0.5]{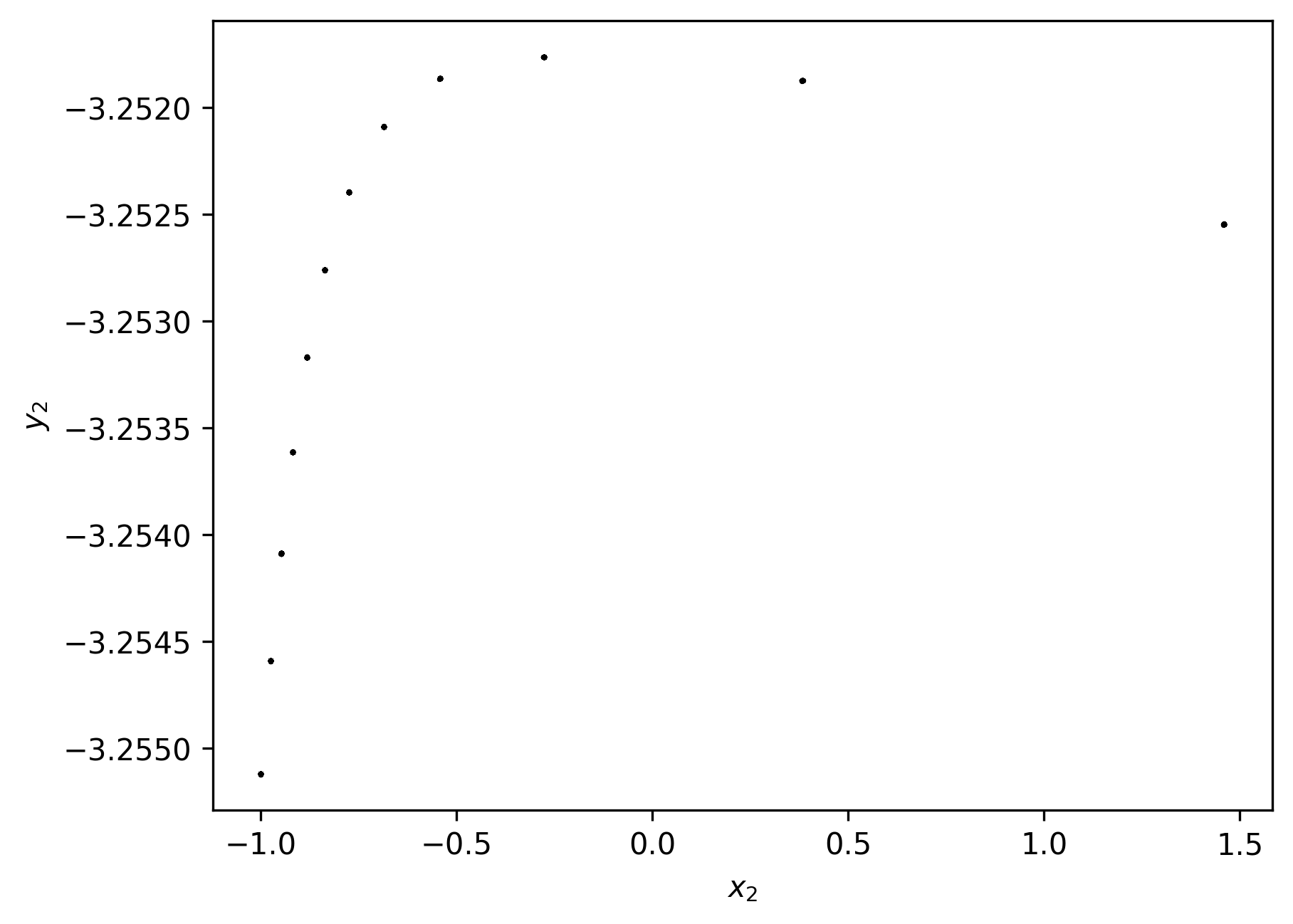}
        \caption{Projection onto the $(x_2,y_2)$ plane}
        \label{fig:nonchaotic_rulkov1_attractor_neuron2}
    \end{subfigure}
    \caption{Projections of the four-dimensional nonchaotic spiking attractor of the asymmetrically coupled Rulkov neuron system}
    \label{fig:nonchaotic_rulkov1_attractor}
    \vspace{0.25cm}
\end{figure*}

\begin{figure*}[ht!]
    \centering
    \begin{subfigure}[t]{0.95\columnwidth}
        \centering
        \includegraphics[scale=0.5]{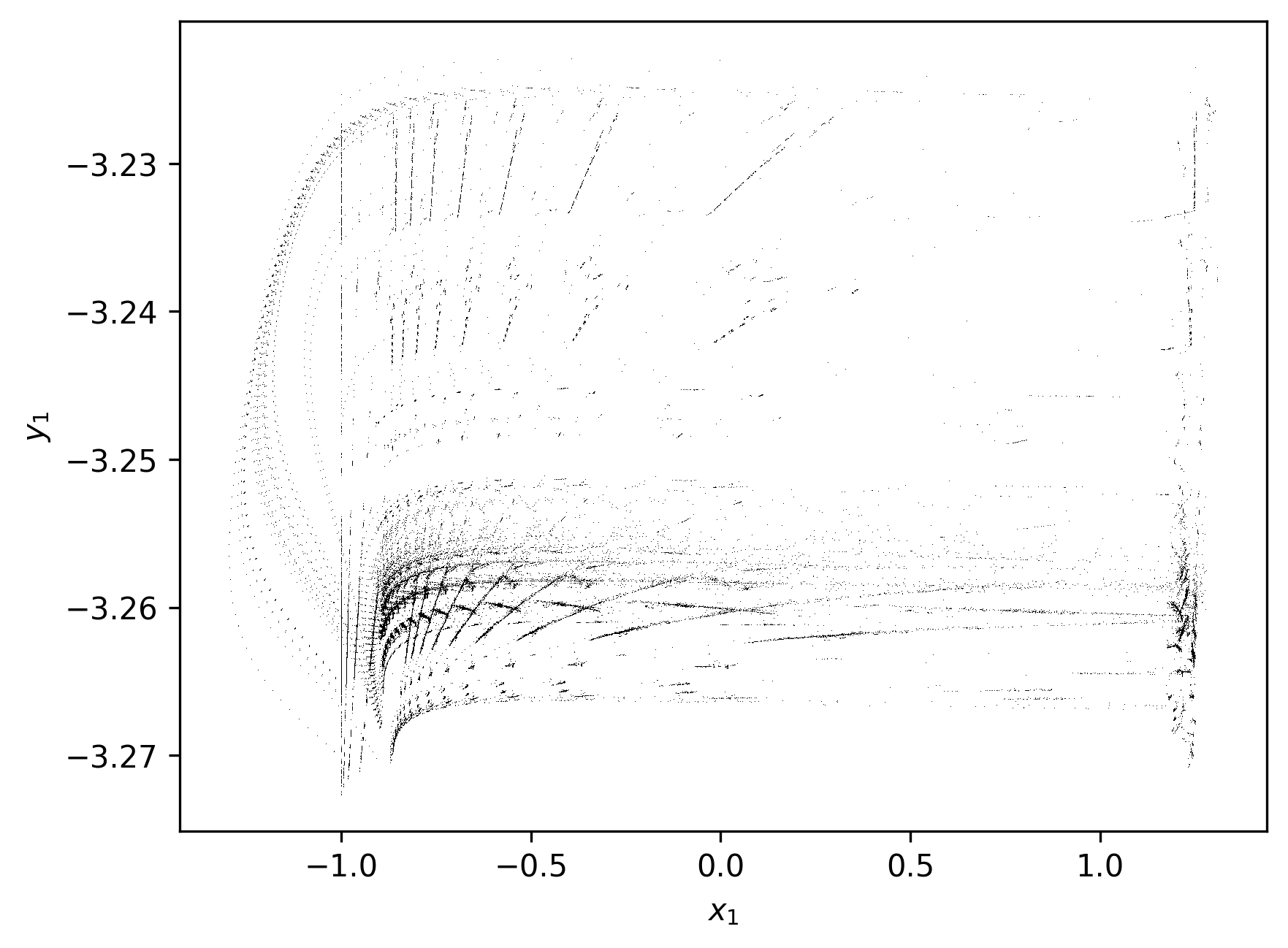}
        \caption{Projection onto the $(x_1,y_1)$ plane}
        \label{fig:chaotic_rulkov1_attractor_neuron1}
        \vspace{0.1cm}
    \end{subfigure}
    \begin{subfigure}[t]{0.95\columnwidth}
        \centering
        \includegraphics[scale=0.5]{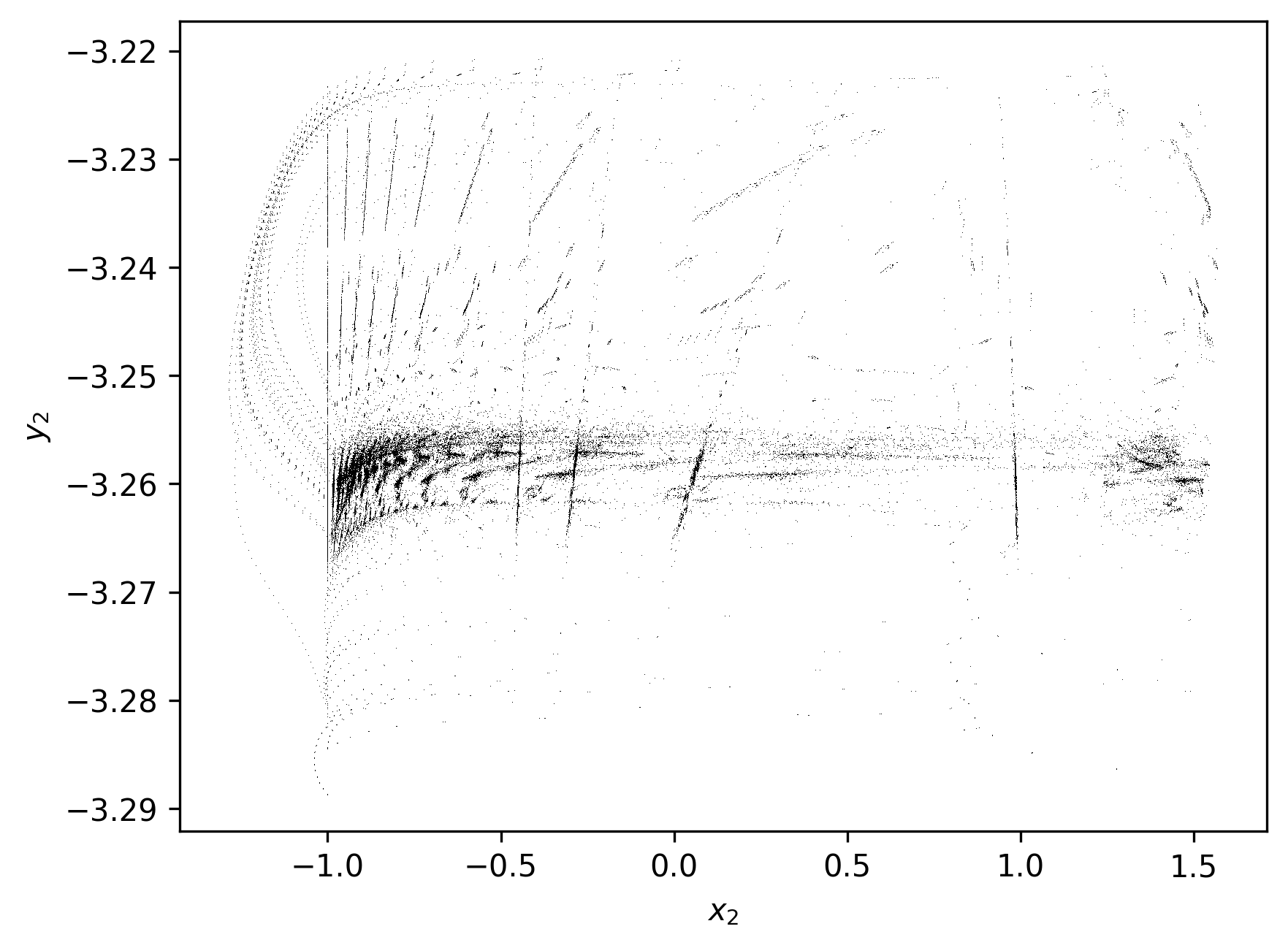}
        \caption{Projection onto the $(x_2,y_2)$ plane}
        \label{fig:chaotic_rulkov1_attractor_neuron2}
    \end{subfigure}
    \caption{Projections of the four-dimensional chaotic spiking-bursting pseudo-attractor of the asymmetrically coupled Rulkov neuron system}
    \label{fig:chaotic_rulkov1_attractor}
\end{figure*}

Although Fig. \ref{fig:asym_coup_rulkov_1_graphs} indicates that this system has two distinct attractors, a nonchaotic spiking attractor and a chaotic spiking-bursting attractor, further analysis of this system shows that this is not completely accurate. To see this, let us examine two initially chaotic orbits of this system. The first orbit we look at is the orbit from Fig. \ref{fig:asym_coup_rulkov_1_chaos}, which has the initial state $\mathbf{X}_{0}=(-0.56, -3.25, -1, -3.25)$. The second orbit is generated from the initial state $\mathbf{X}_{0} = (0.35, -3.25, -1.23, -3.25)$. In Fig. \ref{fig:asym_coup_rulkov_1_long_graphs}, we graph 150,000 iterations of the fast variable orbits emerging from these two initial states. These graphs show that the initial states yield initially chaotic fast variable orbits, but the voltages eventually begin to spike periodically once they get attracted to the nonchaotic spiking attractor. This attraction is made especially clear in Fig. 2 when we zoom into the region where the orbits transition from chaotic to nonchaotic. We propose that these dynamics occur because, during the period when the neurons are exhibiting chaotic spiking-bursting, their slow and fast variables happen to align by chance, which causes their voltages to latch onto each other's spiking and propels the dynamics of the system to the nonchaotic spiking attractor. If this is true, then this system does not have a true chaotic attractor because, by chance, all orbits will eventually get attracted to the nonchaotic spiking attractor. This fact is also verified by our numerical simulations.

Therefore, we conclude that this asymmetrically coupled neuron system isn't multistable. However, this system clearly does have something that is reminiscent of true multistability, with two coexisting qualitatively different behaviors that the orbits of different initial states exhibit. Indeed, we can utilize many of the techniques commonly used to analyze multistability in dynamical systems in the context of this neuron system. Therefore, we will call the phenomenon that this system exhibits ``quasimultistability,'' which represents a multistability that is observed only in the short term. In this system, numerical simulation suggests that quasimultistability is present for orbits of length on the order of $10^4$ (see Fig. \ref{fig:asym_coup_rulkov_1_long_graphs}). We suspect that this type of quasimultistability can occur only in coupled systems similar to this: a small number of coupled identical dynamical systems that exhibit periodic, nonchaotic behavior individually. In these specific systems, the coupling between individually nonchaotic systems can result in chaotic dynamics, and due to there being only a small number of identical systems, they can fall into synchronized periodic orbits in a finite amount of time. For our asymmetrically coupled neuron system, we will treat the short-term chaotic dynamics as occurring on a chaotic ``pseudo-attractor'' on which some orbits end up before being attracted to the nonchaotic spiking attractor. Despite the fact that the spiking attractor is the only true attractor of the system, we find that treating this chaotic spiking-bursting pseudo-attractor as an ``attractor'' as well results in rich geometries in state space emerging from this quasimultistability.

\section{Geometry of the chaotic pseudo-attractor}
\label{sec:attractor}

We now present the results from our analysis of the nonchaotic spiking attractor and the chaotic spiking-bursting pseudo-attractor's geometry. In Figs. \ref{fig:nonchaotic_rulkov1_attractor} and \ref{fig:chaotic_rulkov1_attractor}, we present visualizations of these two attractors by plotting a subset of the orbit of the initial states $\mathbf{X}_{0}=(-0.54, -3.25, -1, -3.25)$ for the nonchaotic attractor and $\mathbf{X}_{0}=(-0.56, -3.25, -1, -3.25)$ for the chaotic pseudo-attractor. In order to visualize these attractors embedded in four-dimensional space $(x_1,y_1,x_2,y_2)$, we plot their projections onto the $(x_1,y_1)$ and $(x_2,y_2)$ planes. It is immediately clear that the spiking attractor (Fig. \ref{fig:nonchaotic_rulkov1_attractor}) is associated with a nonchaotic periodic orbit and the pseudo-attractor (Fig. \ref{fig:chaotic_rulkov1_attractor}) is associated with complex chaotic orbits.

Due to the apparent geometrical complexity of the chaotic pseudo-attractor in Fig. \ref{fig:chaotic_rulkov1_attractor}, we suspect that the pseudo-attractor is fractal and strange. However, the fact that orbits on the pseudo-attractor are chaotic does not guarantee that the pseudo-attractor is fractal, as there exist chaotic attractors that are not strange \cite{grebogi}. The nonchaotic Rulkov map lends itself especially to the possibility of a chaotic attractor that isn't strange due to its resetting mechanism that maps all sufficiently high $x$ values back to $-1$ (Eq. \eqref{eq:rulkov_1_fast_equation}), which always results in at least one Lyapunov exponent of $-\infty$ \cite{bn}. Therefore, we perform a direct computation of the fractal dimension of the chaotic pseudo-attractor using the standard box-counting algorithm \cite{ott}. Using four-dimensional boxes of size $\epsilon=1/20,1/30,1/40$, we get that the box-counting dimension of the chaotic pseudo-attractor is $d\approx 1.84$ with an $R^2$ value of 0.9999, indicating that the chaotic pseudo-attractor is indeed fractal and strange. However, the pseudo-attractor's dimension is still quite small for a geometric object embedded in four-dimensional space, which we attribute to the nonchaotic silence between bursts of spikes [see Fig. \ref{fig:asym_coup_rulkov_1_chaos} and the left sides of Figs. \ref{fig:chaotic_rulkov1_attractor_neuron1} and \ref{fig:chaotic_rulkov1_attractor_neuron2}], as well as the aforementioned resetting mechanism.

\begin{figure*}[ht!]
    \centering
    \begin{subfigure}[t]{0.475\textwidth}
        \centering
        \includegraphics[scale=0.4]{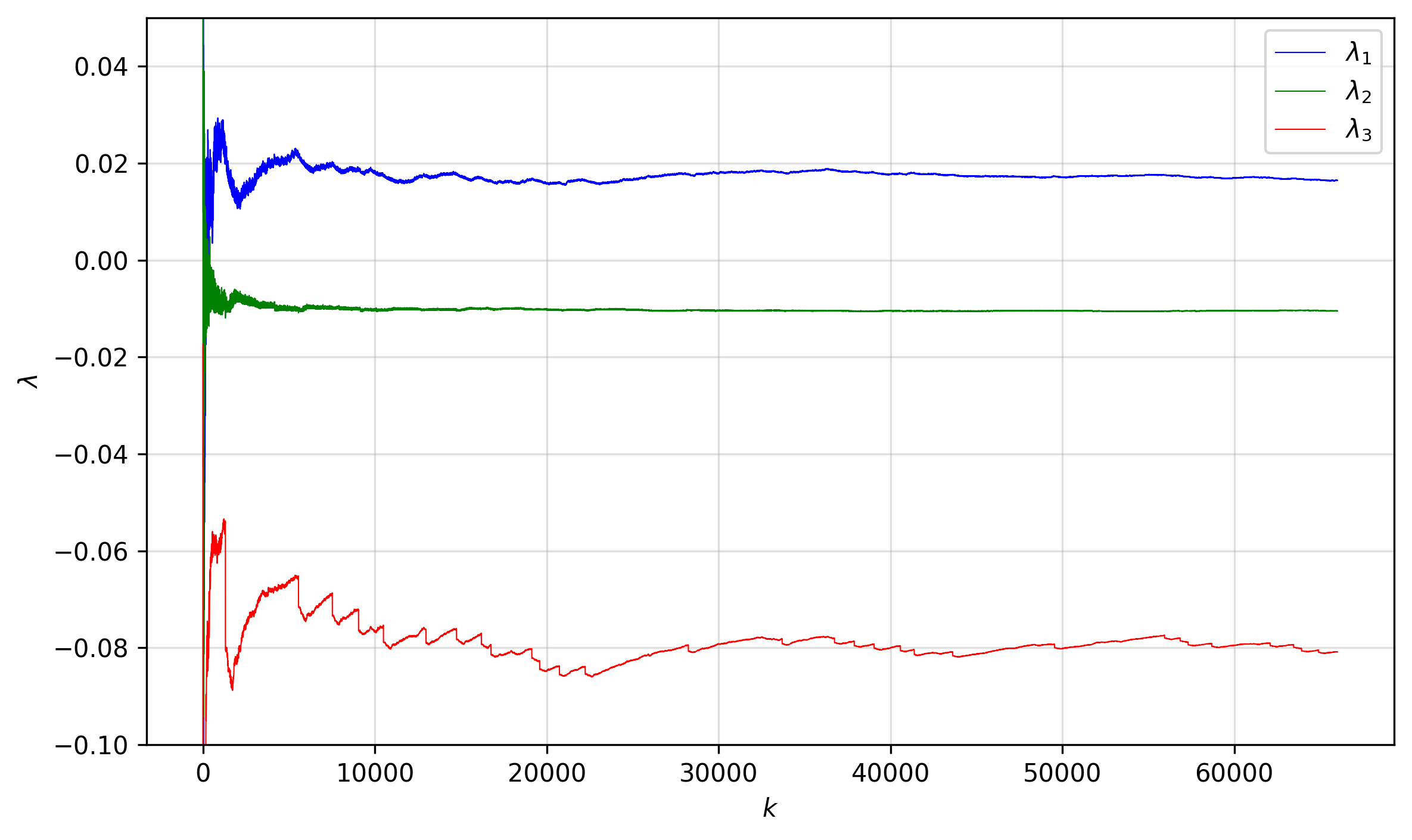}
        \caption{$\lambda_1$, $\lambda_2$, and $\lambda_3$ vs. $k$}
        \label{fig:lyap_conv_all}
        \vspace{0.1cm}
    \end{subfigure}
    \begin{subfigure}[t]{0.475\textwidth}
        \centering
        \includegraphics[scale=0.4]{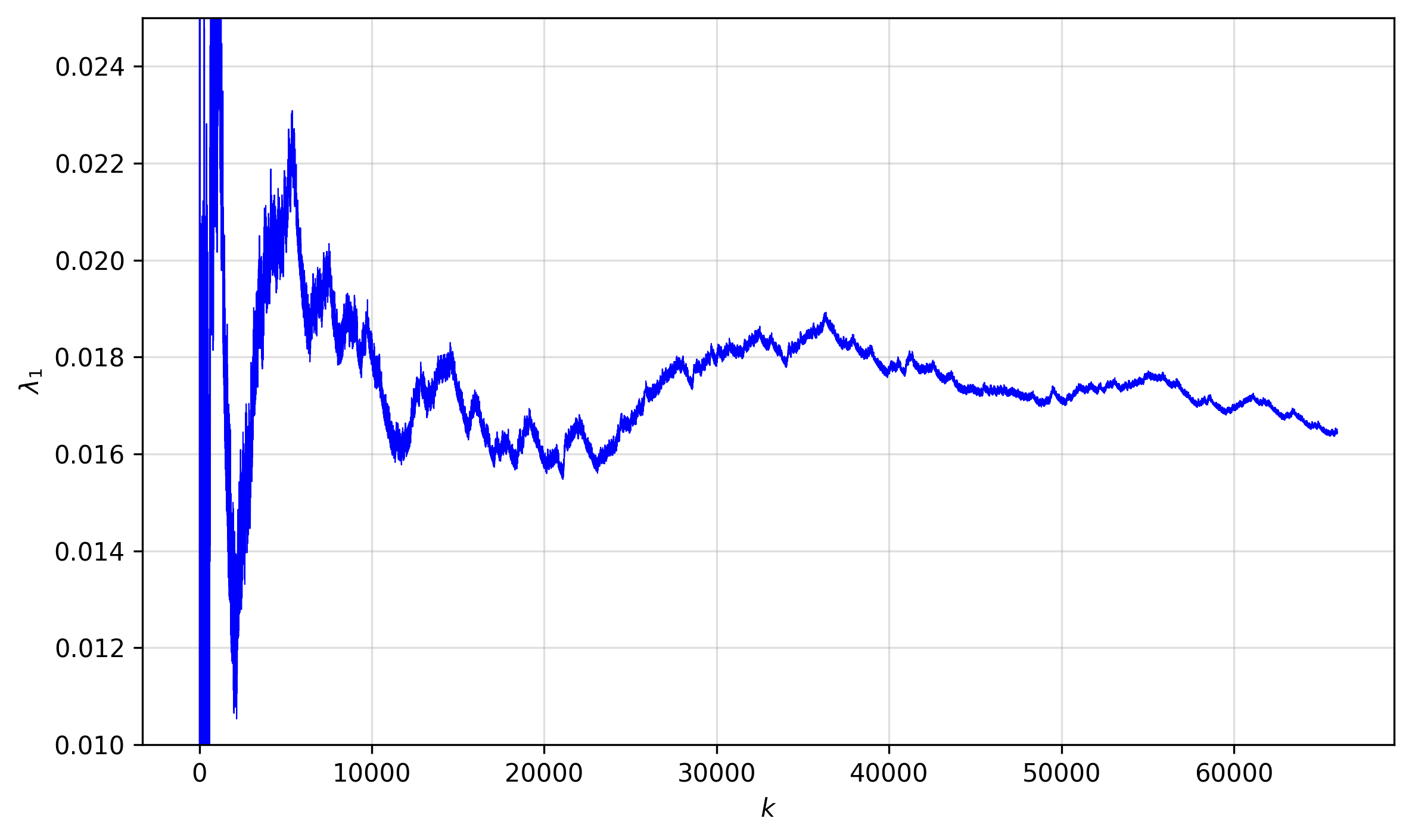}
        \caption{$\lambda_1$ vs. $k$}
        \label{fig:lyap_conv_1}
    \end{subfigure}
    \caption{Plots of the first three Lyapunov exponents at each point in the orbit of $\mathbf{X}_{0} = (-0.56, -3.25, -1, -3.25)$}
    \label{fig:lyap_conv}
\end{figure*}

We are also interested in the Lyapunov dimension
\begin{equation}
    d_l = \kappa + \frac{1}{|\lambda_{\kappa+1}|}\sum_{i=1}^{\kappa}\lambda_i
    \label{eq:dl}
\end{equation}
of the chaotic pseudo-attractor, where $\kappa$ is the largest index such that
\begin{equation}
    \sum_{i=1}^{\kappa}\lambda_i\geq 0,
    \label{eq:kappa}
\end{equation}
and we assume the Lyapunov exponents are ordered from greatest to least \cite{kaplan-yorke}. The Kaplan-Yorke conjecture states that the Lyapunov dimension of typical attractors is equal to their fractal dimension \cite{nichols}. For the chaotic pseudo-attractor orbit displayed in Fig. \ref{fig:chaotic_rulkov1_attractor}, if we naively calculate the Lyapunov dimension using the Lyapunov spectrum at one point in the orbit of an initial state that gets attracted to the chaotic pseudo-attractor, we will most likely not get something that coincides with the pseudo-attractor's true fractal dimension. For example, if we consider the orbit generated from the initial state $\mathbf{X}_{0} = (-0.56, -3.25, -1, -3.25)$, the Lyapunov spectrum is approximately $\lambda\approx\{0.0165,-0.0105,-0.0810,-\infty\}$ at iteration $k=66000$, which is right before the state gets attracted to the nonchaotic spiking attractor [see Fig. \ref{fig:asym_coup_rulkov_1_long_1}]. Using this, Eq. \eqref{eq:dl} indicates that the Lyapunov dimension of the chaotic pseudo-attractor is approximately $d_l\approx 2.07$, which is reasonably far from the pseudo-attractor's true fractal dimension. The reason for this discrepancy lies in the fact that these chaotic dynamics are occurring on a pseudo-attractor, not a true attractor. Although the Kaplan-Yorke conjecture almost always holds for true attractors \cite{farmer}, the fact that we are dealing with a pseudo-attractor means that states do not spend enough time on the pseudo-attractor for their Lyapunov spectrums to capture its true geometry. 

This fact can be easily seen in Fig. \ref{fig:lyap_conv_all}, where we plot the values of the first three Lyapunov exponents at all the iterations of the orbit of the initial state $\mathbf{X}_{0} = (-0.56, -3.25, -1, -3.25)$ that take place on the chaotic pseudo-attractor. Although the state evolves on the pseudo-attractor for tens of thousands of iterations, the figure shows that the Lyapunov exponents have not yet had time to converge to capture the true geometry of the pseudo-attractor, and we see many significant fluctuations in the $\lambda$ values, especially with $\lambda_3$. Of course, it is clear from Eqs. \eqref{eq:dl} and \eqref{eq:kappa} that $\lambda_3$ doesn't contribute the Lyapunov dimension, but Fig. \ref{fig:lyap_conv_1} demonstrates that the maximal Lyapunov exponent $\lambda_1$ has also not converged; it is not oscillating around a mean value, but instead undergoing significant upwards and downwards trends. This explains this rare case where the fractal dimension and Lyapunov dimension of an attractor do not align. It also indicates that our computed Lyapunov exponents can at best be treated as a rough approximation for the true Lyapunov exponents of an orbit. Indeed, in the following sections, we use Lyapunov exponents mainly to determine whether an orbit is chaotic or nonchaotic, not as an exact quantifier for chaotic dynamics.

\section{Basin classification}
\label{sec:classification}

We have established that a quasimultistability exists in this asymmetrically coupled neuron system, so we are now interested in analyzing the geometrical properties of the basins of attraction of the nonchaotic spiking attractor and strange pseudo-attractor. We begin by classifying the basins of attraction using the method of Sprott and Xiong \cite{sprott}, which we briefly outline here. Let $A = \{\mb{a}_1,\mb{a}_2,\hdots\}$ be an attractor \footnote{We assume $A$ is composed of an infinite number of points, but this method can be trivially altered for attractors of a finite number of points.}. Define the mean or ``center of mass'' of $A$ to be
\begin{equation}
    \langle A\rangle = \lim_{j\to\infty}\frac{1}{j}\sum_{i=1}^j\mathbf{a}_i,
    \label{eq:mean}
\end{equation}
and define the standard deviation $\sigma_A$ of $A$ to be
\begin{equation}
    \sigma_A = \sqrt{\lim_{j\to\infty}\frac{1}{j}\sum_{i=1}^j|\mathbf{a}_i-\langle A\rangle|^2}.
    \label{eq:standard-dev}
\end{equation}
Using this, Sprott and Xiong define a ``normalized distance'' $\xi$ from $A$:
\begin{equation}
    \xi = \frac{|\mathbf{x} - \langle A\rangle|}{\sigma_A},
    \label{eq:normalized-distance}
\end{equation}
which allows for the quantification of basins relative to the size of their associated attractors. Now, let $S(\xi)$ be the set of all points that lie in an open $n$-dimensional ball of radius $\xi$ centered at $\langle A\rangle$:
\begin{equation}
    S(\xi) = \{\mathbf{x}:|\mathbf{x}-\langle A\rangle|<\xi\},
\end{equation}
and let $\hat{A}(\xi)$ denote the subset of the basin of attraction $\hat{A}$ contained in the $n$-ball centered at $\langle A\rangle$:
\begin{equation}
    \hat{A}(\xi) = \hat{A}\cap S(\xi).
    \label{eq:a-hat-xi-def}
\end{equation}
Now, we define $P(\xi)$ to be the probability that an initial state $\mathbf{x}_0\in S(\xi)$ is also the basin of attraction of $A$. In other words, 
\begin{equation}
    P(\xi) = \frac{m(\hat{A}(\xi))}{m(S(\xi))},
    \label{eq:p-xi-def}
\end{equation}
where $m(\cdot)$ denotes the measure of a set. In the limit $\xi\to\infty$, $P(\xi)$ follows a power law \cite{sprott}:
\begin{equation}
    P(\xi) = \frac{P_0}{\xi^{\gamma}},
    \label{eq:probability-function-power-law}
\end{equation}
where $P_0$ and $\gamma$ are parameters. Based on the values of $P_0$ and $\gamma$, basins of attraction can be classified from largest to smallest as follows:
\begin{enumerate}
    \item Class 1 basins have $P_0=1$ and $\gamma=0$. These basins occupy all of state space (except perhaps a set of finite measure).
    \item Class 2 basins have $P_0<1$ and $\gamma=0$. These basins occupy a fixed fraction of state space.
    \item Class 3 basins have $0<\gamma<n$, where $n$ is the dimension of the system's state space. These basins extend to infinity in some directions but occupy increasingly small fractions of state space further out.
    \item Class 4 basins have $\gamma=n$. These basins occupy a finite region of state space and have a well-defined relative size $\xi_0 = P_0^{1/n}$.
\end{enumerate}
It is worth noting that we use a shell method for the numerical classification of basins that circumvents the problem of potentially small $P(\xi)$ values for large $\xi$ \cite{bn}. Specifically, defining $\Delta P(2^{k})$ to be the probability that a point lying in the shell centered at $\langle A\rangle$ with inner radius $\xi=2^k$ and outer radius $\xi=2^{k+1}$ is also in the basin of attraction $\hat{A}$, we can iteratively find values of $P(\xi)$ using 
\begin{equation}
    P(2^{k+1}) = \frac{P(2^k)}{2^n} + \left(1-\frac{1}{2^n}\right)\Delta P(2^k).
    \label{eq:iteration-of-shell-method}
\end{equation}

\begin{figure*}[ht!]
    \centering
    \begin{subfigure}{0.545\textwidth}
        \centering
        \includegraphics[scale=0.775]{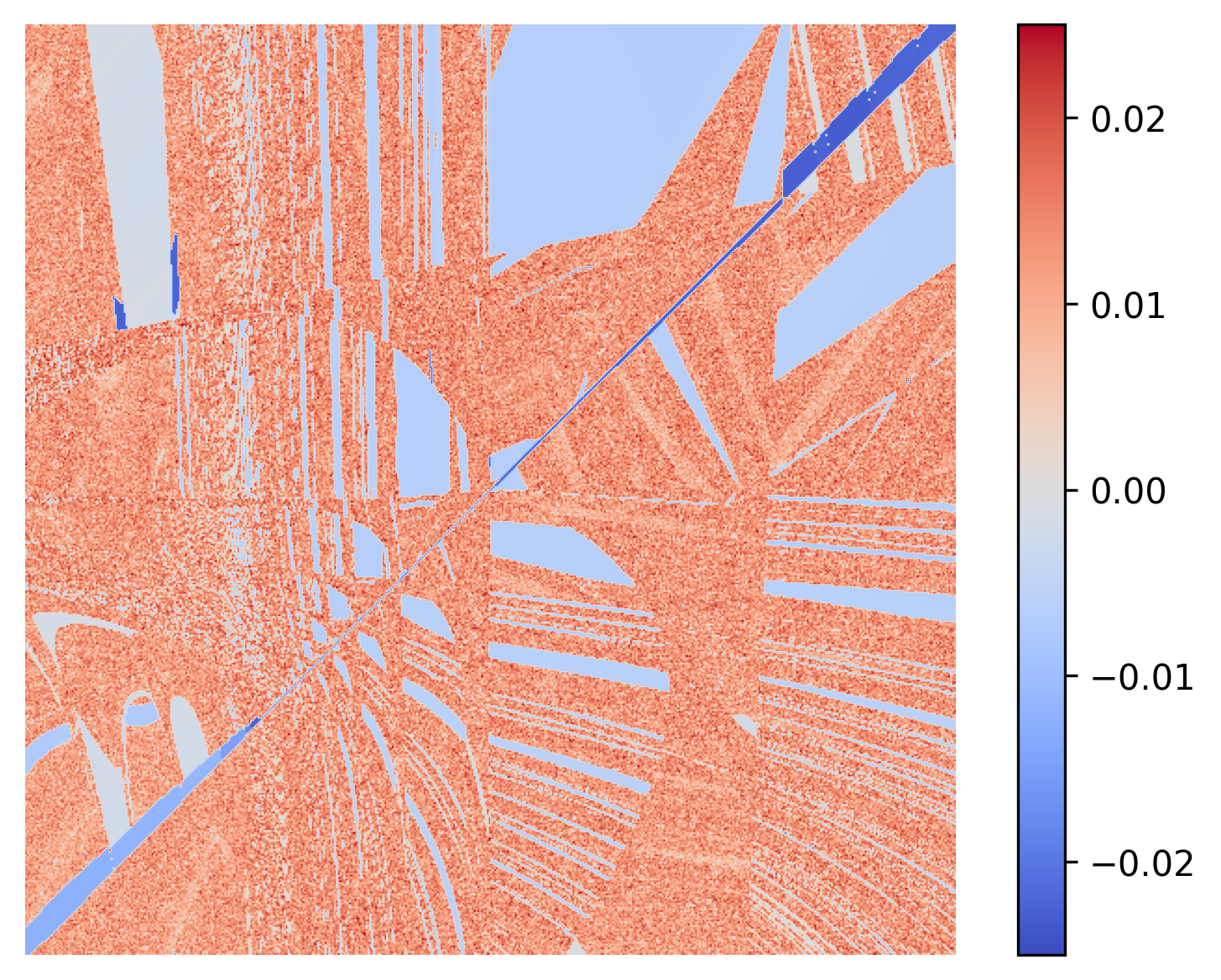}
        \caption{Maximal Lyapunov exponent color map}
        \label{fig:asym_basin_color}
    \end{subfigure}
    \begin{subfigure}{0.445\textwidth}
        \centering
        \includegraphics[scale=0.775]{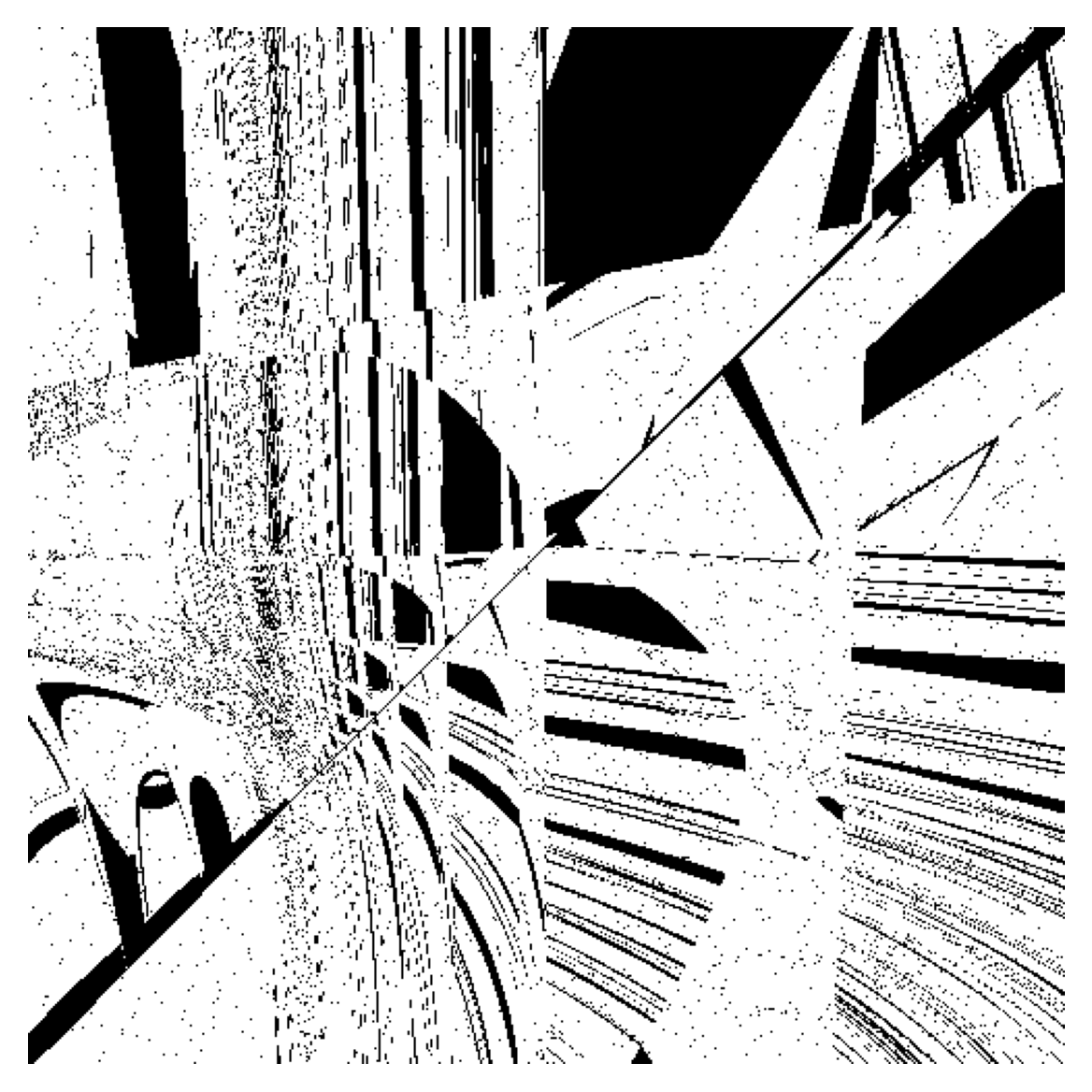}
        \caption{Chaotic (white) and nonchaotic (black) basins}
        \label{fig:asym_basin_bw}
    \end{subfigure}
    \caption{A slice of the basins of attraction of the asymmetrically coupled Rulkov neuron system with $-2\leq x_{1,0}\leq 2$ on the horizontal axis, $-2\leq x_{2,0}\leq 2$ on the vertical axis, and fixed initial $y$ values $y_{1,0}=y_{2,0}=-3.25$}
    \label{fig:asym_basin_graphs}
\end{figure*}

\subsection{Two-dimensional basin slices}
\label{sec:2d-slice}

To analyze the basins of the asymmetrically coupled neuron system, we will first consider a two-dimensional slice of the basins, namely, the intersection of the four-dimensional basins with the set of state space 
\begin{multline}
    S_2'= \{\mathbf{X}: -2<x_1<2,y_1=-3.25,\\
    -2<x_2<2,y_2=-3.25\}.
    \label{eq:s2'}
\end{multline}
We now need a way to determine which attractor an initial state $\mb{X}_0\in S_2'$ is initially attracted to. This is nontrivial because the nonchaotic spiking attractor and chaotic pseudo-attractor occupy similar regions of state space (see Figs. \ref{fig:nonchaotic_rulkov1_attractor} and \ref{fig:chaotic_rulkov1_attractor}). Figure \ref{fig:asym_coup_rulkov_1_long_graphs} illustrates that states attracted to the chaotic pseudo-attractor spend on the order of $10^4$ iterations in the pseudo-attractor before going to the spiking attractor. We find that testing other initial states yields similar results. Therefore, we use an indirect method of determining which basin an initial state is in, saying that an initial state that has a positive maximal Lyapunov exponent after 5,000 iterations is attracted to the chaotic pseudo-attractor, while an initial state that has a negative $\lambda_1$ after the same amount of time is attracted to the spiking attractor. 

Using this method on a large number of initial states $\mb{X}_0\in S_2'$, we present our maximal Lyapunov exponent results with a color map in Fig. \ref{fig:asym_basin_color}. An immediate observation is the beautiful complexity of the different Lyapunov exponents in state space. This complexity is further emphasized when we color the basins properly in Fig. \ref{fig:asym_basin_bw}, with the basin of the chaotic pseudo-attractor (positive $\lambda_1$) shown in white and the basin of the nonchaotic attractor (negative $\lambda_1$) shown in black. Here, we can see that most of $S_2'$ is taken up by the basin of the white chaotic pseudo-attractor, with some black regions, curves, and scattered points of stability scattered throughout. The most obvious of these features is the black line that goes across the diagonal of Fig. \ref{fig:asym_basin_bw} from the bottom left to the top right. This line and its surrounding area are associated with initial states where the two individual neurons' states are the same or very close to being the same. In this case, it makes sense that the neurons will immediately be synchronized, spiking together in the nonchaotic attractor.

Now that we have visualized the basins in $S_2'$, we want to classify the basin slices in the two-dimensional plane containing $S_2'$, which we denote as $S_2$:
\begin{equation}
    S_2 = \{\mathbf{X}: x_1\in\mathbb{R},y_1=-3.25,x_2\in\mathbb{R},y_2=-3.25\}.
\end{equation}
Because $S_2$ is two-dimensional, we also need to specify what is meant by the mean and standard deviation of the four-dimensional attractors within the two-dimensional plane $S_2$. Let us denote a generic four-dimensional attractor as $A_4$ and its associated basin as $\hat{A}_4$ \footnote{In the asymmetrically coupled Rulkov neuron system, $A_4$ could denote either the nonchaotic spiking attractor or the chaotic pseudo-attractor.}. From Eq. \eqref{eq:mean}, if 
\begin{equation}
    A_4=\{\mathbf{a}_1,\mathbf{a}_2,\hdots\}=\left\{\begin{pmatrix}
        a_1\e{1} \\[4pt]
        a_1\e{2} \\[4pt]
        a_1\e{3} \\[4pt]
        a_1\e{4}
    \end{pmatrix},
    \begin{pmatrix}
        a_2\e{1} \\[4pt]
        a_2\e{2} \\[4pt]
        a_2\e{3} \\[4pt]
        a_2\e{4}
    \end{pmatrix},\hdots
    \right\},
\end{equation}
then the true mean of $A_4$ is
\begin{equation}
    \langle A_4\rangle = \lim_{j\to\infty}\frac{1}{j}\sum_{i=1}^j\mathbf{a}_i = \begin{pmatrix}
        \langle x_1 \rangle \\[4pt]
        \langle y_1 \rangle \\[4pt]
        \langle x_2 \rangle \\[4pt]
        \langle y_2 \rangle 
    \end{pmatrix}.
    \label{eq:avgA4}
\end{equation}
We define the ``effective mean'' $\langle A_2\rangle$ of the four-dimensional attractor $A_4$ to be the two-dimensional vector composed of the first and third entries of $\langle A_4\rangle$:
\begin{equation}
    \langle A_2\rangle = \lim_{j\to\infty}\frac{1}{j}\sum_{i=1}^j\begin{pmatrix}
        a_i\e{1} \\[4pt]
        a_i\e{3} 
    \end{pmatrix} = \begin{pmatrix}
        \langle x_1 \rangle \\[4pt]
        \langle x_2 \rangle 
    \end{pmatrix},
\end{equation}
which is the projection of $\langle A_4\rangle$ onto the $(x_1,x_2)$ plane. Similarly, the true standard deviation of $A_4$ in four-dimensional state space is, from Eq. \eqref{eq:standard-dev},
\begin{equation}
    \sigma_{A4} = \sqrt{\lim_{j\to\infty}\frac{1}{j}\sum_{i=1}^j|\mathbf{a}_i-\langle A_4\rangle|^2},
    \label{eq:sigmaA4}
\end{equation}
and we define the ``effective standard deviation'' $\sigma_{A2}$ of $A_4$ to be
\begin{equation}
    \sigma_{A2} = \sqrt{\lim_{j\to\infty}\frac{1}{j}\sum_{i=1}^j\left|\begin{pmatrix}
        a_i\e{1} \\[4pt]
        a_i\e{3} 
    \end{pmatrix}-\langle A_2\rangle\right|^2}.
\end{equation}
Following Eq. \eqref{eq:normalized-distance}, for our classification of the basin slices $\hat{A}_4\cap S_2$, we will consider the normalized two-dimensional distance $\xi_2$ from $\langle A_2\rangle$ of some $\mb{X} = (x_1,-3.25,x_2,-3.25)\in S_2$ to be, 
\begin{equation}
    \xi_2 = \frac{1}{\sigma_{A2}}\left|\begin{pmatrix}
        x_1 \\ x_2
    \end{pmatrix} - \langle A_2\rangle\right|.
\end{equation}

To classify the basin slices of the chaotic pseudo-attractor and nonchaotic spiking attractor, we want to find the two probability functions $P_w(\xi_2)$ and $P_b(\xi_2)$ associated with the white (chaotic) and black (nonchaotic) basins, respectively, which we accomplish using an iterative Monte Carlo method. Specifically, we first randomly select states from the open disk of radius $\xi_2=1$ centered at $\langle A_2\rangle$ and calculate the maximal Lyapunov exponents of the initial states' orbits to determine which basin they end up in. Using this, we calculate the value of $P(1)$. Then, starting from $k=0$, we repeat the same process for the two-dimensional shell centered at $\langle A_2\rangle$ with inner radius $\xi_2=2^k$ and outer radius $\xi_2=2^{k+1}$ to calculate the value of $\Delta P(2^k)$. Then, we use Eq. \eqref{eq:iteration-of-shell-method} to find $P(2^{k+1})$ and repeat the process for $k+1$. One important numerical note is that, given we are testing values of $\xi_2=2^k$, the maximum value of $k$ must be chosen with care. On the one hand, since we are interested in the limit $\xi\to\infty$, $k$ must be large enough that we capture the geometry of the basins far away from the attractors. On the other hand, since we are only iterating 5000 times, it also must be small enough that initial states go to an attractor fast enough to get an accurate Lyapunov exponent calculation on it. With a computer experiment, we determine that a good maximum value of $k$ is 11.

\begin{table}[t]
    \centering
    \begin{tabular}{c|c|c}
        $\xi_2$ & $P_w(\xi_2)$ & $P_b(\xi_2)$ \\
        \hline \\ [-10px]
        $2^6=64$ & 0.8365 & 0.1530 \\
        $2^7=128$ & 0.8736 & 0.1237 \\
        $2^8=256$ & 0.8919 & 0.1010 \\
        $2^9=512$ & 0.9148 & 0.0839 \\
        $2^{10}=1024$ & 0.9419 & 0.0675 \\
        $2^{11}=2048$ & 0.9360 & 0.0544 \\
    \end{tabular}
    \caption{Some approximate $P(\xi_2)$ values of the asymmetrically coupled Rulkov neuron system's chaotic (white) and nonchaotic (black) basins}
    \label{tab:p_function_asym_2_values}
\end{table}

In Table \ref{tab:p_function_asym_2_values}, we display our numerical results for the values of $P_w(\xi_2)$ and $P_b(\xi_2)$, where we neglect values of $k$ from 0 to 5 since we are interested in large values of $\xi_2$. We can immediately see that $P_w(\xi_2)$ is behaving a bit oddly, as Eq. \eqref{eq:probability-function-power-law} indicates that $P(\xi)$ functions should either decrease or stay the same. For this reason, we first turn our attention to $P_b(\xi_2)$, which is behaving as expected. Performing a linear regression calculation on $\log_2 P_b(\xi_2)$ against $\log_2\xi_2$, we get that
\begin{equation}
    \log_2 P_b(\xi_2) = -0.2957\log_2\xi_2 - 0.9360.
    \label{eq:asym-basin-2-lin-reg}
\end{equation}
Since $\gamma\approx0.2957$ is between 0 and $n=2$, we can conclude that the basin slice of the nonchaotic spiking attractor is Class 3, meaning it extends to infinity in some directions but takes up increasingly small fractions of state space as we move outwards. This makes sense because the farther away an initial state is from the nonchaotic spiking attractor, the lower the chance there will be for the neurons to immediately synchronize with each other. In the limit $\xi_2\to\infty$, we expect that the only initial conditions that get attracted to the spiking attractor immediately are the ones where $x_{1,0}=x_{2,0}$, which is the diagonal in Fig. \ref{fig:asym_basin_graphs}. For this reason, we can conclude that the black basin slice in $S_2$ has finite measure since the line $x_{1,0}=x_{2,0}$ doesn't contribute to the measure of the two-dimensional black basin slice. 

To classify the white basin, we use the fact that
\begin{equation}
    P_w(\xi_2) + P_b(\xi_2) = 1
    \label{eq:sum-the-basins}
\end{equation}
by the definitions of the spiking attractor and chaotic pseudo-attractor: if an initial state does not immediately fall into a nonchaotic spiking orbit with negative $\lambda_1$, then it goes to a chaotic orbit with positive $\lambda_1$. This fact can be confirmed by adding the data horizontally across Table \ref{tab:p_function_asym_2_values}, which gives values very close to 1. Therefore, even though $P_w(2^{11})$ is less than $P_w(2^{10})$ due to the aforementioned difficulty of calculating Lyapunov exponents for orbits of initial states far away from their attractors, the values of $P_w(\xi_2)$ are indeed approaching 1. This can be seen by exponentiating both sides of Eq. \eqref{eq:asym-basin-2-lin-reg}:
\begin{equation}
    P_b(\xi_2) \approx \frac{0.5227}{\xi_2^{0.2957}},
\end{equation}
which goes to 0 as $\xi_2\to\infty$. Therefore, because we are interested in this very limit, the basin slice of the chaotic pseudo-attractor must be Class 1. Although it does not take up all of $S_2$, it does take up all of $S_2$ barring a set of finite measure, which is the black basin. If a random initial state were selected from the entirety of $S_2$, there would indeed be a 100\% probability of it being attracted to the chaotic pseudo-attractor, as the white basin slice has an infinite measure, while its complement has a finite measure.

\subsection{Four-dimensional basins}
\label{sec:4d-basins}

To classify the entire basins living in all of four-dimensional space $S_4 = \mathbb{R}^4$, it is evident that we should define normalized distance in this set as 
\begin{equation}
    \xi_4 = \frac{|\mathbf{X}-\langle A_4\rangle|}{\sigma_{A4}},
\end{equation}
using the definitions from Eqs. \eqref{eq:avgA4} and \eqref{eq:sigmaA4}. This is similar to the process for the two-dimensional slices, with a few noteworthy changes. One nontrivial aspect of this is choosing a random state $\mb{X}_0$ from a four-dimensional open ball with radius $\xi_4$ centered at $\langle A_4\rangle$. In two-dimensional space with an open disk $|\mb{x}-\langle A_2\rangle|<\xi_2$, we simply choose a random $r\in[0,\xi_2)$ and a random $\phi\in[0, 2\pi)$, then say the random initial state is $\mb{x}_0 = \langle A_2\rangle + (r\cos\phi, r\sin\phi)$. In four-dimensional space, we can use the analog to spherical coordinates in four-dimensions, where four coordinates $r\in[0,\xi_4)$, $\theta_1\in[0,\pi)$, $\theta_2\in[0,\pi)$, and $\phi\in[0,2\pi)$ are chosen randomly. Then, a random initial state $\mathbf{X}_0\in S(\xi_4)$ is given by \cite{blumenson}
\begin{equation}
    \mathbf{X}_0 = \langle A_4\rangle + \begin{pmatrix}
        r\sin\theta_1\sin\theta_2\cos\phi \\
        r\sin\theta_1\sin\theta_2\sin\phi \\
        r\sin\theta_1\cos\theta_2 \\
        r\cos\theta_1
    \end{pmatrix}.
\end{equation}
Another numerical note is, because $y$ is a slow variable, starting far away from where it will eventually end up (around $y=-3.25$) inevitably results in the initial state taking even longer to reach its attractor. For this reason, we raise the number of test point iterations for Lyapunov exponent calculation from 5000 to 20000, which significantly raises computation time. 

\begin{table}[t]
    \centering
    \begin{tabular}{c|c|c}
        $\xi_4$ & $P_w(\xi_4)$ & $P_b(\xi_4)$ \\
        \hline \\ [-10px]
        $2^0=1$ & 0.8480 & 0.1520 \\
        $2^1=2$ & 0.8649 & 0.1351 \\
        $2^2=4$ & 0.8491 & 0.1509 \\
        $2^3=8$ & 0.8537 & 0.1463 \\
        $2^4=16$ & 0.8690 & 0.1310 \\
        $2^5=32$ & 0.8549 & 0.1451 \\
        $2^6=64$ & 0.8766 & 0.1234 \\
        $2^7=128$ & 0.8760 & 0.1240 \\
        $2^8=256$ & 0.8629 & 0.1371 \\
    \end{tabular}
    \caption{Some approximate $P(\xi_4)$ values of asymmetrically coupled Rulkov neuron system's chaotic (white) and nonchaotic (black) basins}
    \label{tab:p_function_asym_4_values}
\end{table}

\begin{figure*}[htp!]
    \centering
    \begin{subfigure}{0.545\textwidth}
        \centering
        \includegraphics[scale=0.775]{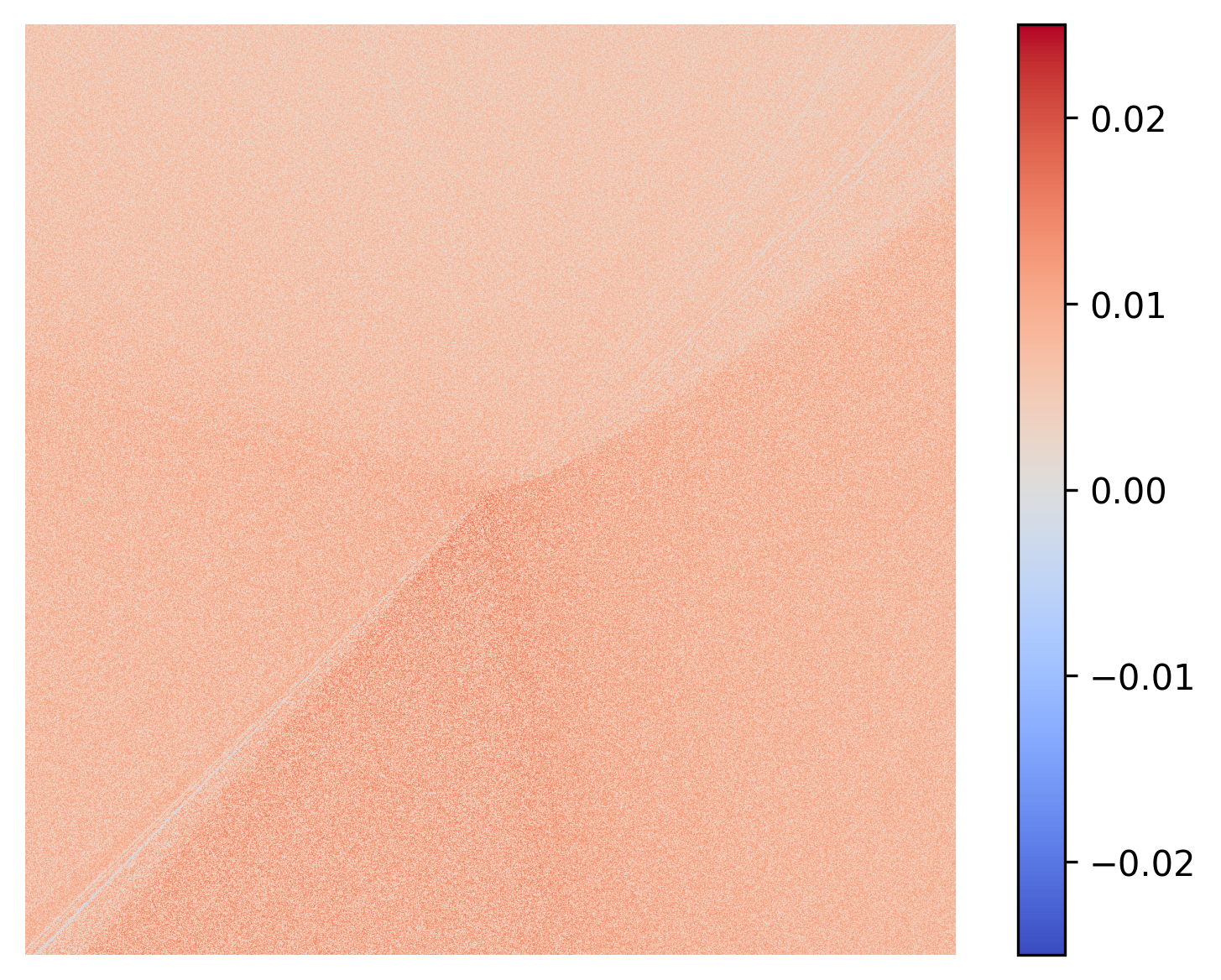}
        \caption{Maximal Lyapunov exponent color map}
        \label{fig:asym_basin_color_y}
    \end{subfigure}
    \begin{subfigure}{0.445\textwidth}
        \centering
        \includegraphics[scale=0.775]{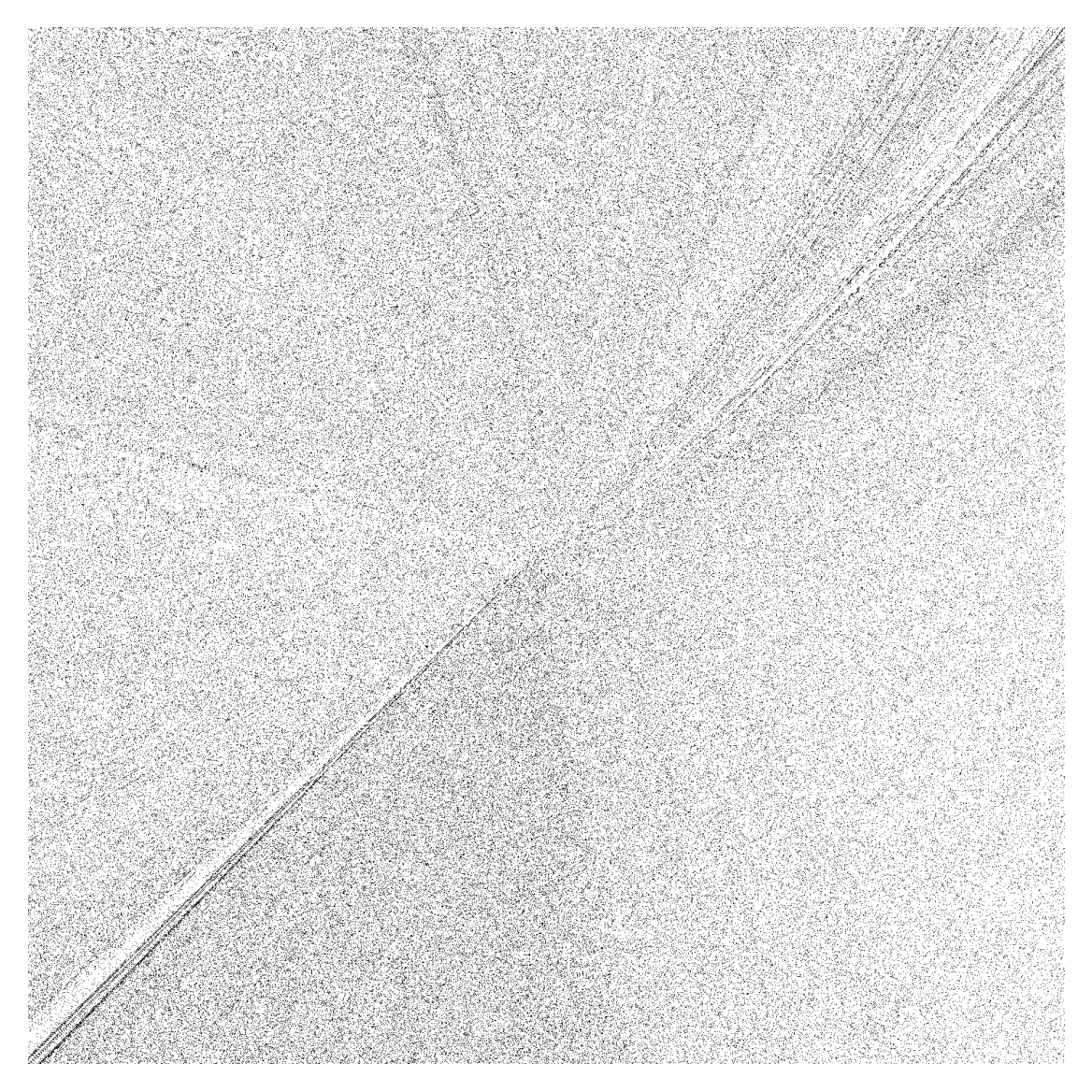}
        \caption{Chaotic (white) and nonchaotic (black) basins}
        \label{fig:asym_basin_bw_y}
    \end{subfigure}
    \caption{A slice of the basins of attraction of the asymmetrically coupled Rulkov neuron system with $-256\leq y_{1,0}\leq 256$ on the horizontal axis, $-256\leq y_{2,0}\leq 256$ on the vertical axis, and fixed initial $x$ values $x_{1,0}=-1$ and $x_{2,0}=1$}
    \label{fig:asym_basin_graphs_y}
\end{figure*}

The results for the probability function values of the white basin $P_w(\xi_4)$ are shown in Table \ref{tab:p_function_asym_4_values}. In the table, the $P_b(\xi_4)$ values are indirectly calculated using $P_b(\xi_4) = 1-P_w(\xi_4)$, which follows from Eq. \eqref{eq:sum-the-basins}. Observing the values in Table \ref{tab:p_function_asym_4_values}, it appears that $P_w(\xi_4)$ and $P_b(\xi_4)$ remain relatively constant as $\xi_4$ is varied, which is a rather unexpected result given our analysis of the two-dimensional basin slices. Running linear regressions on the log-log plots of both $P_w(\xi_4)$ and $P_b(\xi_4)$, we find that 
\begin{equation}
    P_w(\xi_4) = 0.0042\xi_4 - 0.2315
    \label{eq:asym-white-basin-4-lin-reg}
\end{equation}
with $R^2=0.392$, and
\begin{equation}
    P_b(\xi_4) = -0.0260\xi_4 - 2.7540
    \label{eq:asym-black-basin-4-lin-reg}
\end{equation}
with $R^2=0.386$. These low $R^2$ values confirm that there is no upward or downward trend in $P_w(\xi_4)$ or $P_b(\xi_4)$. This means that, within the numerical variability of the Monte Carlo method, the value of the basin classification parameter $\gamma$ is effectively 0 for both the white and black basins. From Eqs. \eqref{eq:asym-white-basin-4-lin-reg} and \eqref{eq:asym-black-basin-4-lin-reg}, the value of the other basin classification parameter $P_0$ is $P_0=0.8517$ for the white basin and $P_0=0.1482$ for the black basin. Both of these are between 0 and 1, so both the white and black basins are Class 2, meaning they both occupy a fixed fraction of four-dimensional state space. This makes sense considering the data in Table \ref{tab:p_function_asym_4_values}, but again, it is an unexpected result. 

Our explanation for why the white and black basins are Class 2 relies on the fact that the nonchaotic Rulkov map is a slow-fast system. Because $x$ is a fast variable and $y$ is a slow variable, $x_1$ and $x_2$ adjust much faster than $y_1$ and $y_2$, so if we consider some initial state $\mathbf{X}_0=(x_{1,0},y_{1,0},x_{2,0},y_{2,0})$ some distance $\xi_4$ away from either attractor $A_4$, $y_1$ and $y_2$ will slowly drift towards $\langle y_1\rangle$ and $\langle y_2\rangle$, respectively, and $x_1$ and $x_2$ will easily keep up with their slow attraction. By the time $y_1$ and $y_2$ reach the region around $\langle y_1\rangle$ and $\langle y_2\rangle$, $x_1$ and $x_2$ will also be near $\langle x_1\rangle$ and $\langle x_2\rangle$, respectively. This will occur regardless of how much farther away $x_{1,0}$ and $x_{2,0}$ are from $\langle x_1\rangle$ and $\langle x_2\rangle$ compared to the distance from $y_{1,0}$ and $y_{2,0}$ to $\langle y_1\rangle$ and $\langle y_2\rangle$, respectively, since the fast variables $x_1$ and $x_2$ evolve so much more quickly than the slow variables $y_1$ and $y_2$. Thus, once $y_1$ and $y_2$ get close to the attractors, we are effectively starting in $S_2'$ (or a square of similar size close and parallel/skew to it). Indeed, comparing the fractions of state space that the white ($P_0=0.8517$) and black ($P_0=0.1482$) basins take up to the values of $P_w(\xi_2)$ and $P_b(\xi_2)$ for the small values of $\xi_2$ in Table \ref{tab:p_function_asym_2_values}, we find that these align quite well.

Although we cannot visualize the entirety of these four-dimensional basins, we can get a grasp of this Class 2 basin behavior by graphing a different two-dimensional slice of the basins, namely, one where $y_{1,0}$ and $y_{2,0}$ are varied but $x_{1,0}$ and $x_{2,0}$ are fixed. In Fig. \ref{fig:asym_basin_graphs_y}, we graph a large range of initial $y$ values $-256\leq y_{1,0}\leq 256$ and $-256\leq y_{2,0}\leq 256$ while fixing the initial $x$ values at $x_{1,0}=-1$ and $x_{2,0}=1$. In Fig. \ref{fig:asym_basin_bw_y}, we can see a dispersed, seemingly random distribution of white and black basin points with no defined pattern in their arrangement besides some subtle outward extending darker and lighter regions. Similarly, in Fig. \ref{fig:asym_basin_color_y}, the maximal Lyapunov exponent also appears random. This apparent randomness makes sense from our discussion of how far-away points evolve toward the attractors. Because of how slow $y_1$ and $y_2$ move towards their eventual values, by the time the state reaches a square close to $S_2'$, the fast variables $x_1$ and $x_2$ have evolved so much that they are essentially randomized. In other words, any difference in $y_1$ and $y_2$ between two initial states will result in substantially different $x_1$ and $x_2$ values once the state reaches a square close to $S_2'$. This essentially turns the initial state from $S_4$ into a random initial state in $S_2'$ (see Fig. \ref{fig:asym_basin_graphs}).

The important observation from Fig. \ref{fig:asym_basin_bw_y} is that despite the angular variances in the distribution of white and black points, the distribution appears to remain the same regardless of the radial distance from the center. Even though the plot spans a large range of different $y$ values, this consistent distribution characteristic of Class 2 basins is present: the fraction of state space Class 2 basins take up doesn't depend on the distance from their attractors. Figure \ref{fig:asym_basin_bw_y} also lacks the black diagonal that appears in Fig. \ref{fig:asym_basin_bw}. This line of stability doesn't appear in this graph because, as previously mentioned, it is defined by
\begin{equation}
    x_{1,0} = y_{1,0} = x_{2,0} = y_{2,0},
\end{equation}
which the slice in Fig. \ref{fig:asym_basin_graphs_y} intersects only at its center.

\begin{figure*}[htp!]
    \centering
    \includegraphics[width=0.75\linewidth]{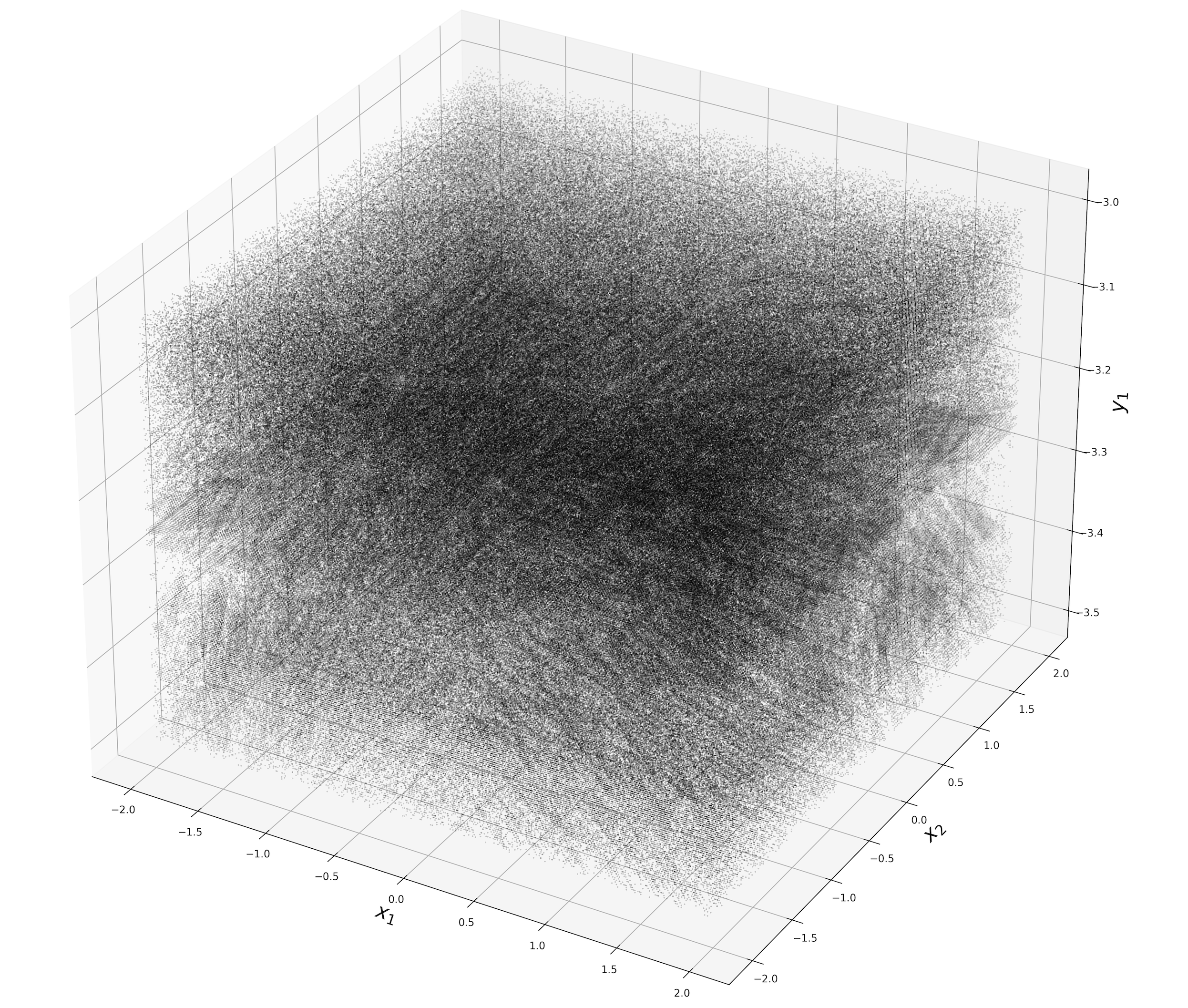}
    \caption{A three-dimensional slice of the black (nonchaotic) basin of the asymmetrically coupled Rulkov neuron system with $-2\leq x_{1,0}\leq 2$, $-2\leq x_{2,0}\leq 2$, $-3.5\leq y_{1,0}\leq -3$, and fixed initial $y_2$ values $y_{2,0}=-3.25$}
    \label{fig:3d-basin}
\end{figure*}

\begin{figure*}[hbp!]
    \centering
    \begin{subfigure}{0.245\textwidth}
        \centering
        \includegraphics[scale=0.35]{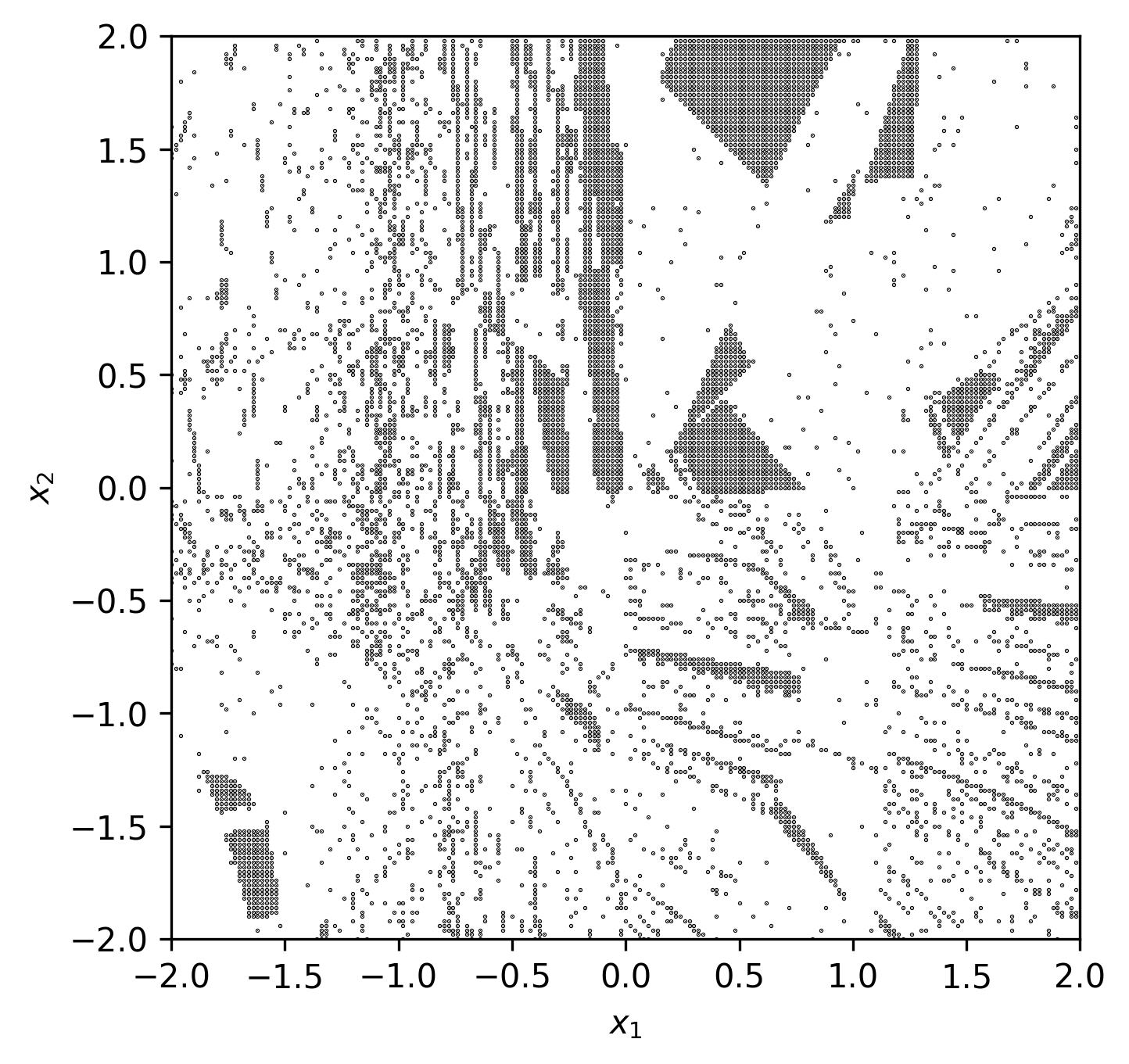}
        \vspace{-0.2cm}
        \caption{$y_{1,0} = -3.24$}
        \label{fig:2d_slice_-3.24}
        \vspace{0.25cm}
    \end{subfigure}
    \begin{subfigure}{0.245\textwidth}
        \centering
        \includegraphics[scale=0.35]{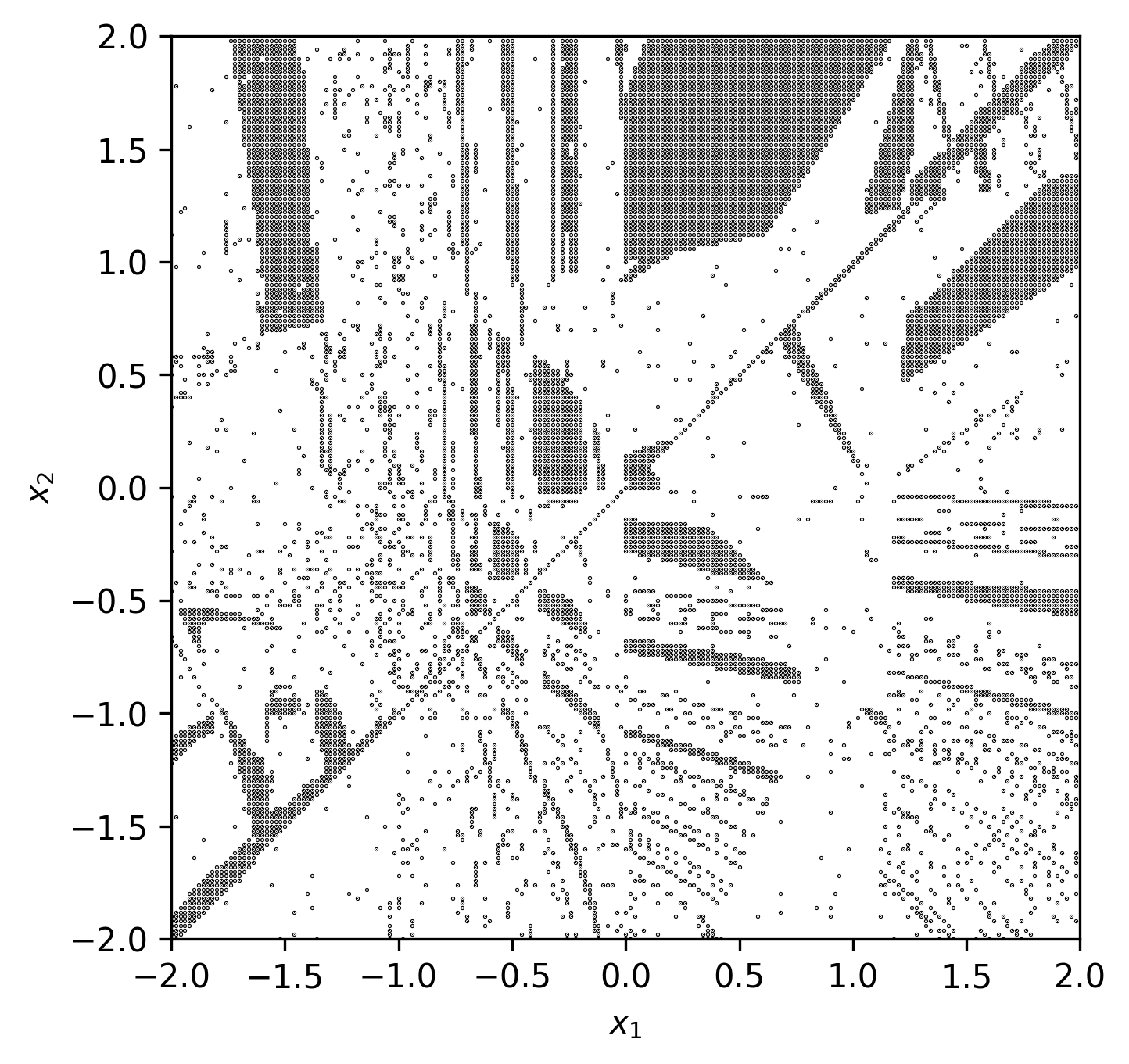}
        \vspace{-0.2cm}
        \caption{$y_{1,0} = -3.25$}
        \label{fig:2d_slice_-3.25}
        \vspace{0.25cm}
    \end{subfigure}
    \begin{subfigure}{0.245\textwidth}
        \centering
        \includegraphics[scale=0.35]{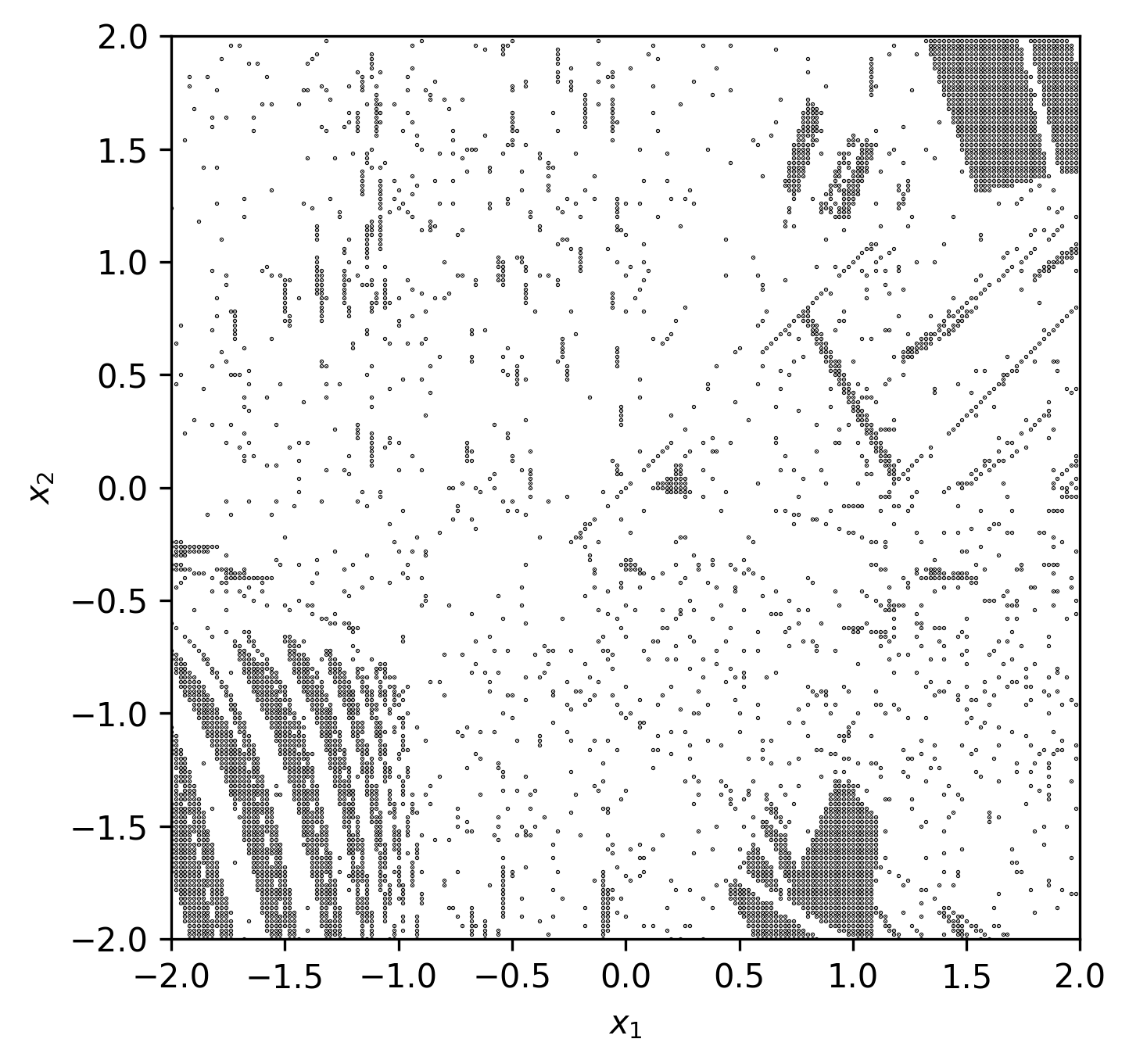}
        \vspace{-0.2cm}
        \caption{$y_{1,0} = -3.26$}
        \label{fig:2d_slice_-3.26}
        \vspace{0.25cm}
    \end{subfigure}
    \begin{subfigure}{0.245\textwidth}
        \centering
        \includegraphics[scale=0.35]{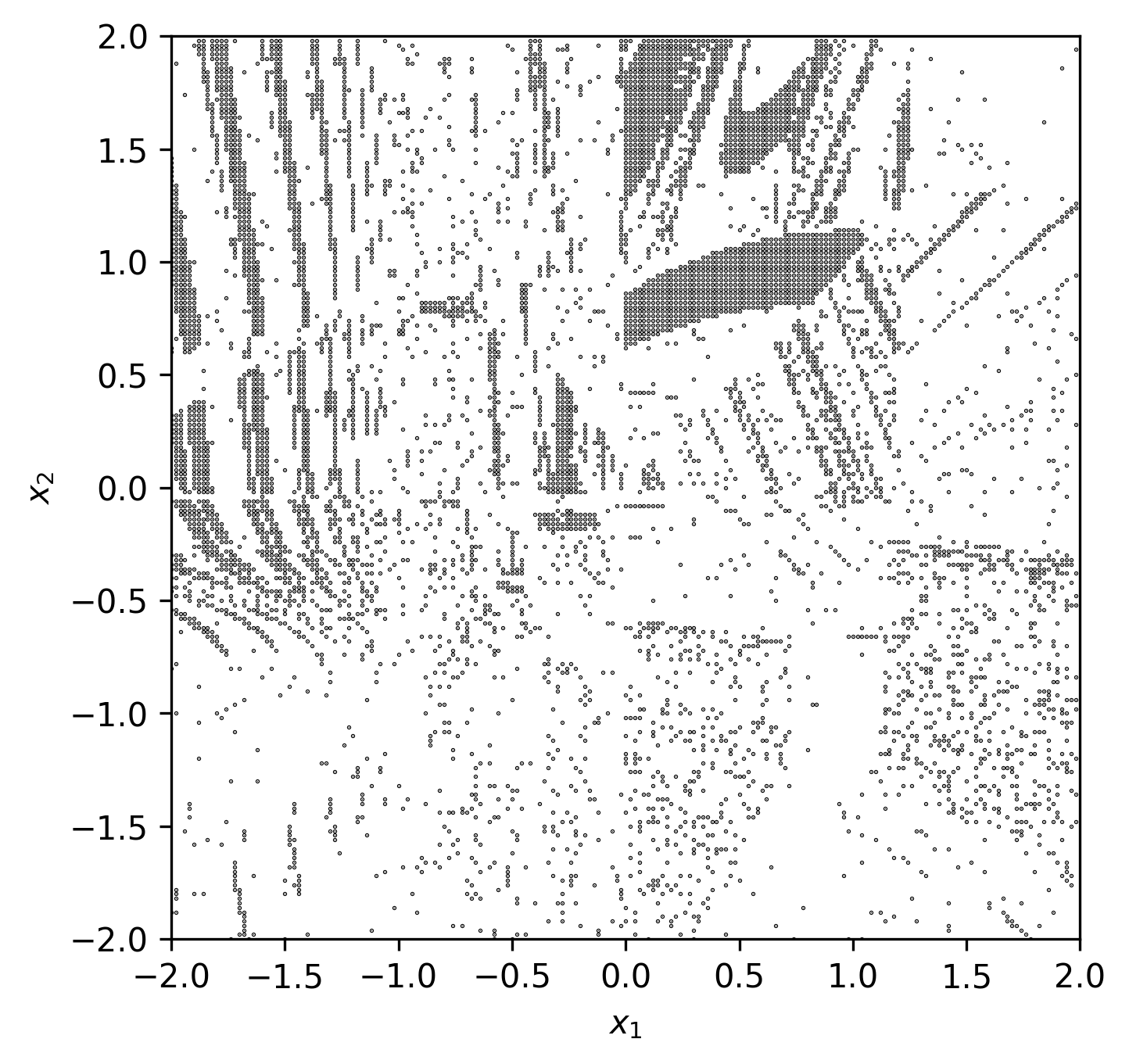}
        \vspace{-0.2cm}
        \caption{$y_{1,0} = -3.27$}
        \label{fig:2d_slice_-3.27}
        \vspace{0.25cm}
    \end{subfigure}
    \begin{subfigure}{0.245\textwidth}
        \centering
        \includegraphics[scale=0.35]{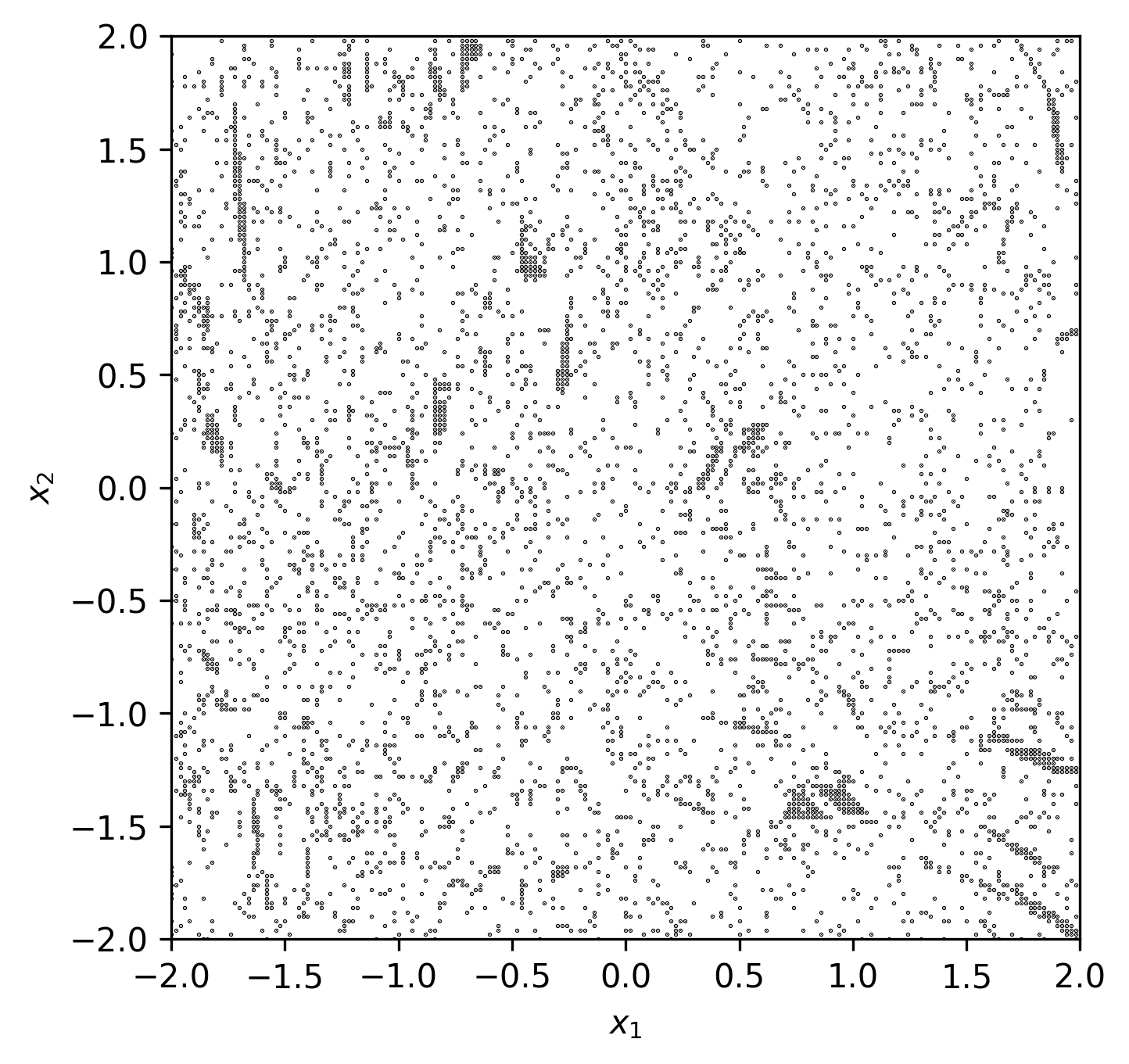}
        \vspace{-0.2cm}
        \caption{$y_{1,0} = -3.15$}
        \label{fig:2d_slice_-3.15}
    \end{subfigure}
    \begin{subfigure}{0.245\textwidth}
        \centering
        \includegraphics[scale=0.35]{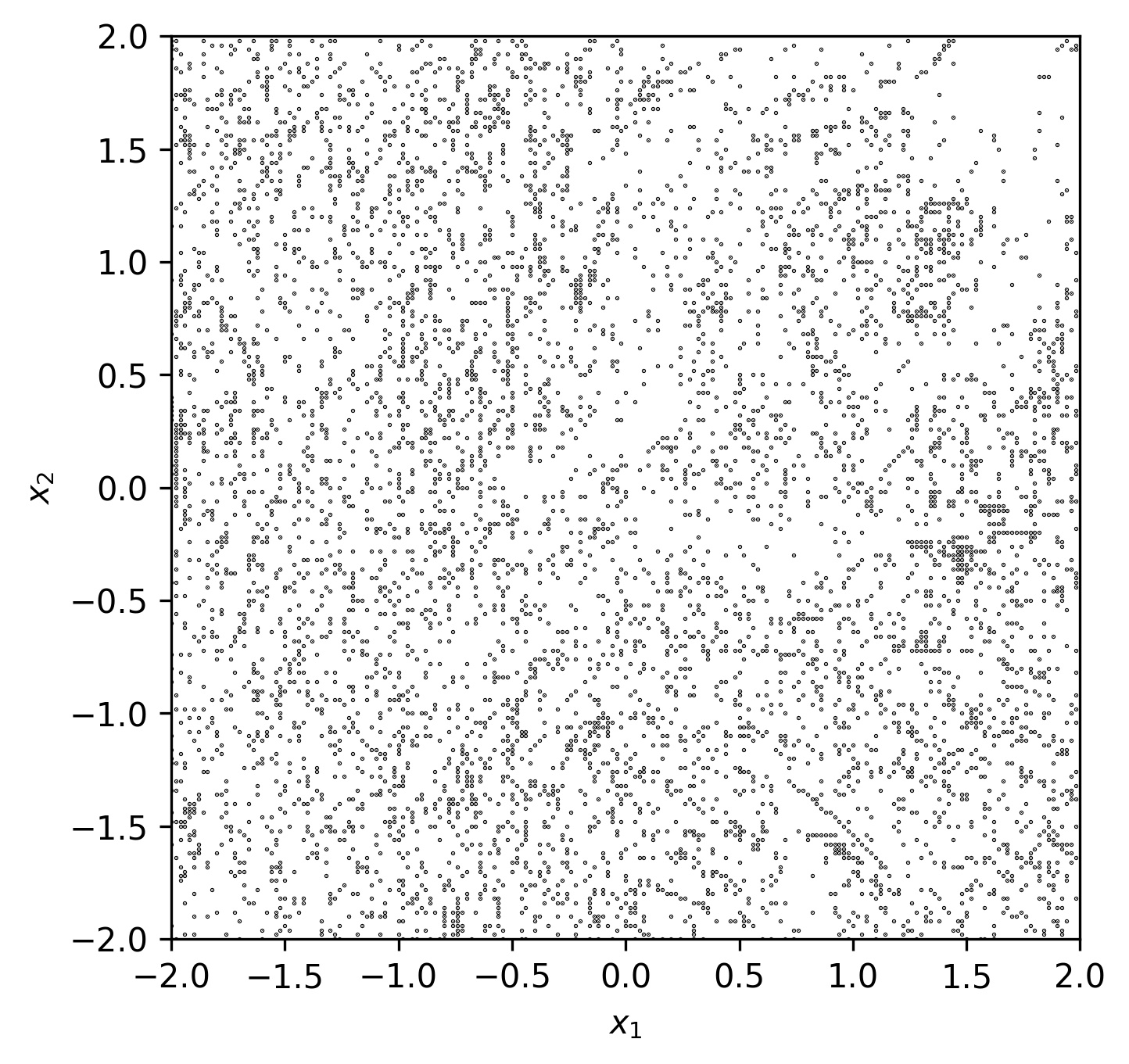}
        \vspace{-0.2cm}
        \caption{$y_{1,0} = -3.2$}
        \label{fig:2d_slice_-3.2}
    \end{subfigure}
    \begin{subfigure}{0.245\textwidth}
        \centering
        \includegraphics[scale=0.35]{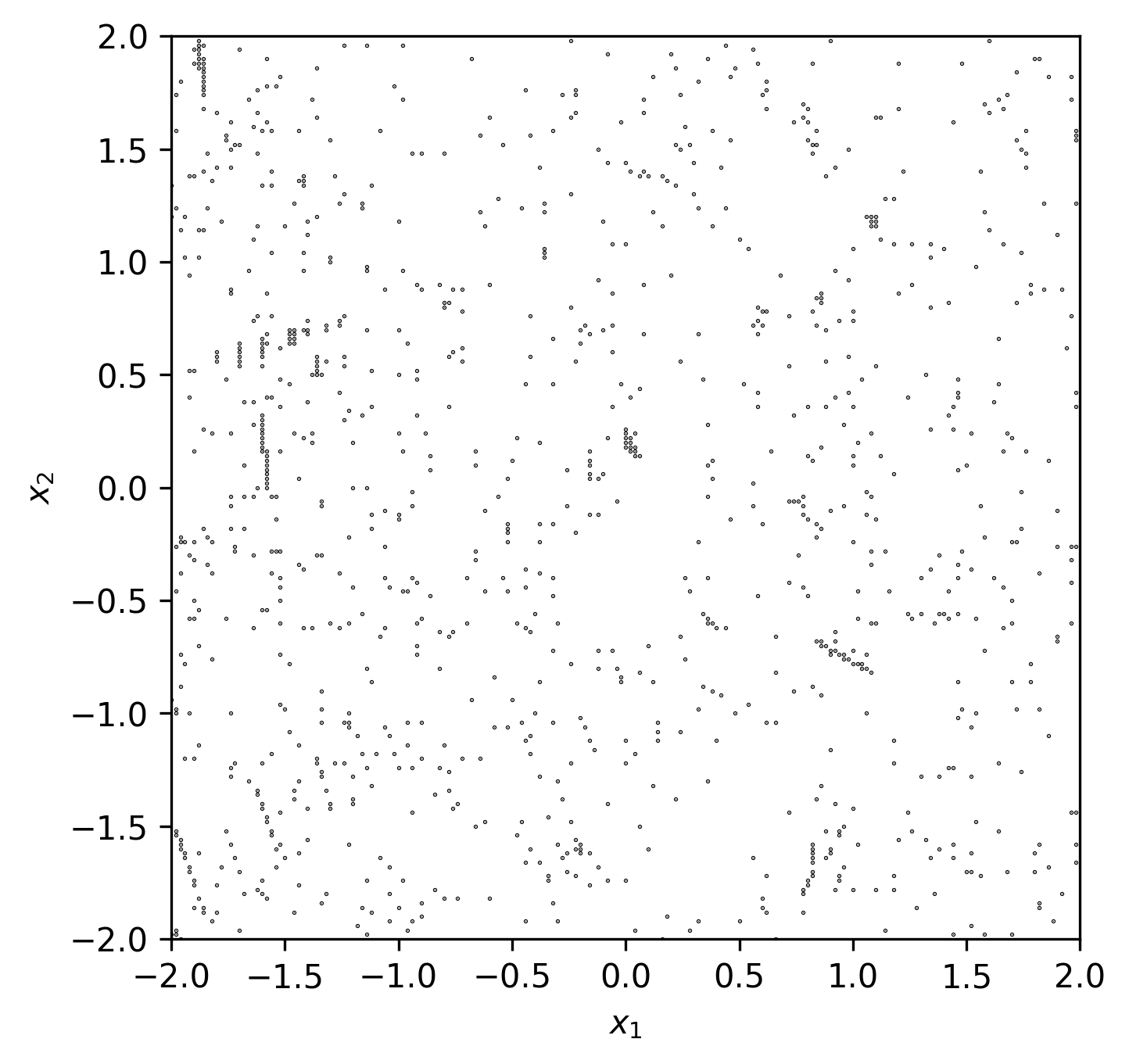}
        \vspace{-0.2cm}
        \caption{$y_{1,0} = -3.3$}
        \label{fig:2d_slice_-3.3}
    \end{subfigure}
    \begin{subfigure}{0.245\textwidth}
        \centering
        \includegraphics[scale=0.35]{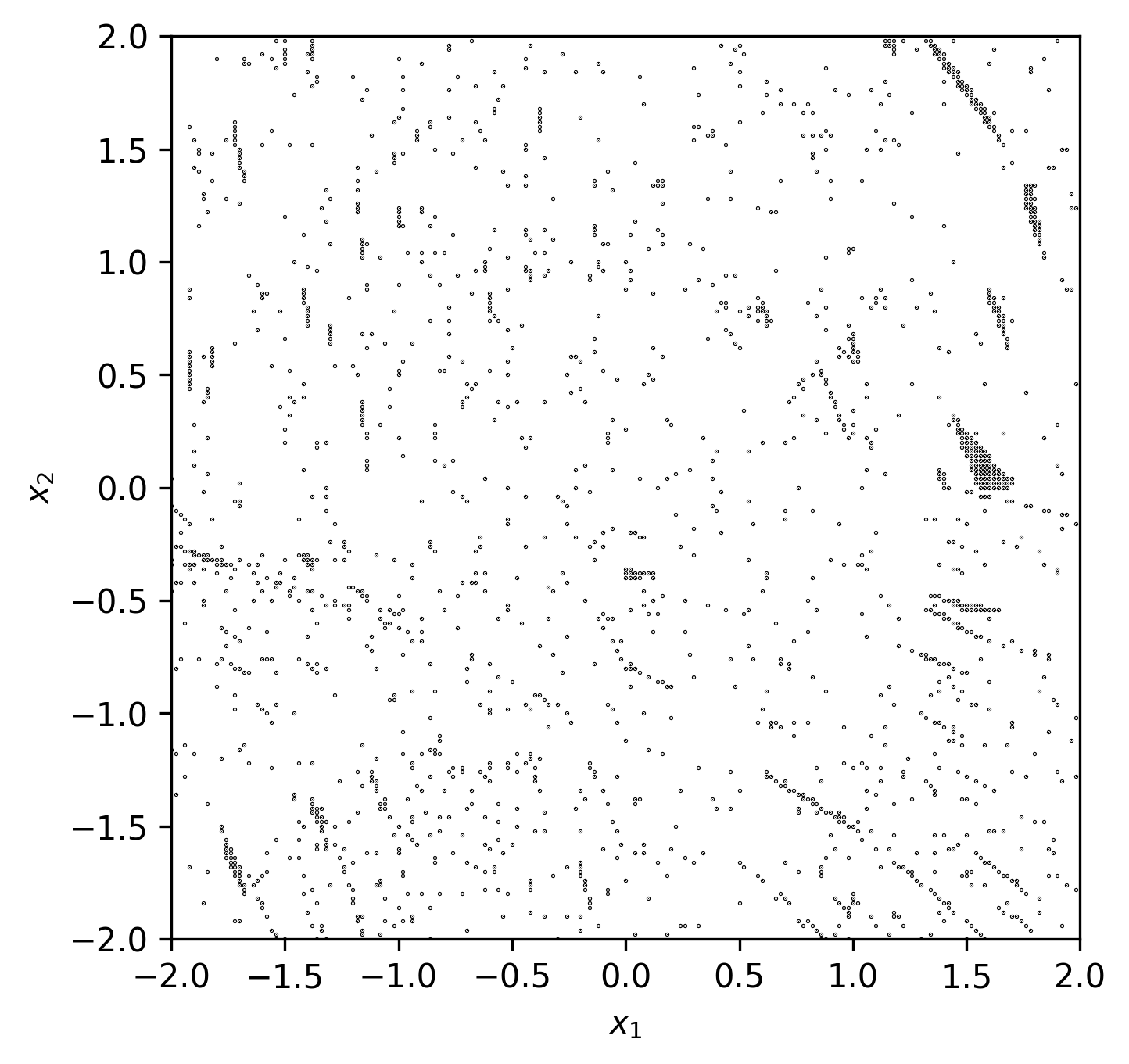}
        \vspace{-0.2cm}
        \caption{$y_{1,0} = -3.35$}
        \label{fig:2d_slice_-3.35}
    \end{subfigure}
    \caption{Select horizontal slices of Fig. \ref{fig:3d-basin}}
    \label{fig:2d_slices}
\end{figure*}

To get a more comprehensive picture of the four-dimensional basins, we take inspiration from Ref. \cite{3d} to present a small three-dimensional slice of the black basin in Fig. \ref{fig:3d-basin}, where we mark initial states that get attracted to the nonchaotic spiking attractor with a small black point (the empty space represents the white basin). The initial states included in this plot have initial $x_1, x_2, y_1$ values between $-2\leq x_{1,0}\leq 2$, $-2\leq x_{2,0}\leq 2$, and $-3.5\leq y_{1,0}\leq -3$, and fixed initial $y_2$ values $y_{2,0}=-3.25$. An immediate observation is the dense complexity of the basins in three-dimensional space, certainly encapsulating the aforementioned seemingly random distribution of basin points without any obvious defined structure. To get a better view of this complex three-dimensional plot, we show eight select $(x_1,x_2)$ plane slices of the three-dimensional basins in Fig. \ref{fig:2d_slices}. To get a sense of the level of detail in Figs. \ref{fig:3d-basin} and \ref{fig:2d_slices}, a good comparison to make is between Figs. \ref{fig:asym_basin_bw} and \ref{fig:2d_slice_-3.25}, which show the same two-dimensional basin slice. An observation from Fig. \ref{fig:3d-basin} is the relative abundance of black basin points around the $y_{1,0} = -3.25$ plane. Figs. \ref{fig:2d_slice_-3.24}--\ref{fig:2d_slice_-3.27} support this, and they also show that more interesting geometrical structures arise in this region as opposed to the random distributions in Figs. \ref{fig:2d_slice_-3.15}--\ref{fig:2d_slice_-3.35}. This observation is to be expected since the region around the $y_{1,0} = -3.25$ plane corresponds to initial states with $y_{1,0}\approx y_{2,0}$, so it makes sense that initial states in this region will have a higher chance of synchronizing and being attracted to the spiking attractor immediately, especially considering that we are examining a small range of initial $x$ values. The more defined geometric structures in Figs. \ref{fig:2d_slice_-3.24}--\ref{fig:2d_slice_-3.27} also make sense from this perspective since the slow variables do not have to travel far to reach their eventual values, which we mentioned earlier is the reason for the apparent random scattering in Figs. \ref{fig:asym_basin_bw_y} and \ref{fig:2d_slice_-3.15}--\ref{fig:2d_slice_-3.35}.


\section{Fractal basin boundary analysis}
\label{sec:boundary}

In Figs. \ref{fig:asym_basin_graphs} and \ref{fig:asym_basin_graphs_y}, it is clear that the basins of the nonchaotic spiking attractor and strange pseudo-attractor are not divided by a smooth boundary. This naturally leads us to suspect that the basin boundary $\Sigma$ between the white and black basins might be fractal, which would lead to uncertainty regarding which attractor a given initial state will end up in \cite{grebogi-final-state}. In this section, we use the uncertainty exponent method \cite{mcdonald} to determine the dimension of the basin boundary $\Sigma$ and the degree to which final states (in the short term) are uncertain. Briefly, the uncertainty exponent method works as follows. Consider a dynamical system in $n$-dimensional state space, and let $\mb{x}_0$ be an initial state with a small uncertainty $\epsilon$, meaning the true initial state may be anywhere in the $n$-ball $|\mb{x}-\mb{x}_0|<\epsilon$. Let $A$ and $C$ be the only two attractors of the dynamical system, and let $\hat{A}$ and $\hat{C}$ be their associated basins of attraction. $\mb{x}_0$ will be attracted to either $A$ or $C$, but the uncertainty $\epsilon$ may cause the perturbed initial state to be in a different basin than $\mb{x}_0$. Then, we expect the probability that we will falsely predict where a random uncertain state in $|\mb{x}-\mb{x}_0|<\epsilon$ will end up, denoted by $\varrho(\epsilon)$, to follow a power law:
\begin{equation}
    \varrho(\epsilon) = \epsilon^{\mathfrak{u}},
\end{equation}
where $\mathfrak{u}\leq 1$ is called the uncertainty exponent. Then, when $\mathfrak{u}$ is small, significantly minimizing initial state uncertainty will have only a small effect on final state uncertainty. It is shown in \cite{mcdonald} that the uncertainty exponent $\mathfrak{u}$ is related to the dimension of the basin boundary dividing $\hat{A}$ and $\hat{C}$ by
\begin{equation}
    \mathfrak{u} = n-d,
    \label{eq:uncertainty-exp-frac-dim}
\end{equation}
which means that a small uncertainty exponent is associated with a highly fractal basin boundary.

Returning to the asymmetrically coupled Rulkov neuron system, we present the results of our analysis of the basin boundary $\Sigma$ between the white and black basins using uncertainty exponents. Similar to how we classified the intersection of the neuron system's basins with $S_2$ and $S_4$, we examine the uncertainty exponents and fractalization of both $\Sigma\cap S_2'$ and $\Sigma\cap S_4'$. We have already defined $S_2'$ as a specific square-shaped subset of $S_2$ (Eq. \eqref{eq:s2'}), and we will similarly define $S_4'\subset S_4$ to be the four-dimensional hypercube
\begin{multline}
    S_4'= \{\mathbf{X}: -2<x_1<2,-1<y_1<-5,\\
    -2<x_2<2,-1<y_2<-5\}.
    \label{eq:s4'}
\end{multline}

First, let us examine the set $\Sigma\cap S_2'$, displayed in Fig. \ref{fig:asym_basin_bw} as the boundary between the white and black basins. We denote the uncertainty exponent of this basin boundary subset as $\mathfrak{u}_2$ and its associated probability function as $\varrho_2(\epsilon)$. To calculate $\varrho_2(\epsilon)$, we pick a random initial state $\mathbf{X}_0\in S_2'$ and test four specific perturbed states
\begin{equation}
    \begin{pmatrix}
        x_{1,0}+\epsilon \\
        -3.25 \\
        x_{2,0} \\
        -3.25
    \end{pmatrix},
    \begin{pmatrix}
        x_{1,0}-\epsilon \\
        -3.25 \\
        x_{2,0} \\
        -3.25
    \end{pmatrix},
    \begin{pmatrix}
        x_{1,0} \\
        -3.25 \\
        x_{2,0}+\epsilon \\
        -3.25
    \end{pmatrix},
    \begin{pmatrix}
        x_{1,0} \\
        -3.25 \\
        x_{2,0}-\epsilon \\
        -3.25
    \end{pmatrix}
\end{equation}
to see if any of them end up in a different basin from $\mathbf{X}_0$.

Calculating the uncertainty exponent of the basin boundary in four-dimensional state space $\Sigma\cap S_4'$ is a natural extension from calculating $\mathfrak{u}_2$. We similarly denote the uncertainty exponent of this four-dimensional basin boundary set as $\mathfrak{u}_4$ and its associated probability function as $\varrho_4(\epsilon)$. The step of picking a random initial state $\mathbf{X}_0\in S_4'$ is much easier than picking a random initial state from a four-dimensional ball (which we did for basin classification) since we chose our region of analysis $S_4'$ to be a hypercube. However, since $S_4'$ is a four-dimensional region, we now need to test eight perturbed states:
\begin{equation}
    \begin{gathered}
        \begin{pmatrix}
            x_{1,0}+\epsilon \\
            y_{1,0} \\
            x_{2,0} \\
            y_{2,0}
        \end{pmatrix},
        \begin{pmatrix}
            x_{1,0}-\epsilon \\
            y_{1,0} \\
            x_{2,0} \\
            y_{2,0}
        \end{pmatrix},
        \begin{pmatrix}
            x_{1,0} \\
            y_{1,0}+\epsilon \\
            x_{2,0} \\
            y_{2,0}
        \end{pmatrix},
        \begin{pmatrix}
            x_{1,0} \\
            y_{1,0}-\epsilon \\
            x_{2,0} \\
            y_{2,0}
        \end{pmatrix}, \\
        \begin{pmatrix}
            x_{1,0} \\
            y_{1,0} \\
            x_{2,0}+\epsilon \\
            y_{2,0}
        \end{pmatrix},
        \begin{pmatrix}
            x_{1,0} \\
            y_{1,0} \\
            x_{2,0}-\epsilon \\
            y_{2,0}
        \end{pmatrix},
        \begin{pmatrix}
            x_{1,0} \\
            y_{1,0} \\
            x_{2,0} \\
            y_{2,0}+\epsilon
        \end{pmatrix},
        \begin{pmatrix}
            x_{1,0} \\
            y_{1,0} \\
            x_{2,0} \\
            y_{2,0}-\epsilon
        \end{pmatrix}.
    \end{gathered}
\end{equation}

\begin{table}[t]
    \centering
    \begin{tabular}{c|c|c}
        $\epsilon$ & $\varrho_2(\epsilon)$ & $\varrho_4(\epsilon)$ \\
        \hline \\ [-10px]
        $2^0=1$ & 0.710 & 0.404 \\
        $2^{-1}=1/2$ & 0.714 & 0.416 \\
        $2^{-2}=1/4$ & 0.658 & 0.394 \\
        $2^{-3}=1/8$ & 0.571 & 0.392 \\
        $2^{-4}=1/16$ & 0.516 & 0.339 \\
        $2^{-5}=1/32$ & 0.409 & 0.349 \\
        $2^{-6}=1/64$ & 0.340 & 0.341 \\
        $2^{-7}=1/128$ & 0.255 & 0.328 \\
        $2^{-8}=1/256$ & 0.219 & 0.323 \\
        $2^{-9}=1/512$ & 0.172 & 0.308 \\
        $2^{-10}=1/1024$ & 0.126 & 0.306 \\
        $2^{-11}=1/2048$ & 0.092 & 0.302 \\
    \end{tabular}
    \caption{Some approximate $\varrho(\epsilon)$ values for the basin boundary sets $\Sigma\cap S_2'$ and $\Sigma\cap S_4'$ of the asymmetrically coupled Rulkov neuron system}
    \label{tab:uncertainty_exp_asym_coup_values}
\end{table}

Because $S_2'$ and $S_4'$ are relatively small and close to the attractors, we can safely determine which basin an initial state belongs to by calculating its Lyapunov spectrum with a 5000-iteration long orbit. The results of our numerical calculations of some values of $\varrho_2(\epsilon)$ and $\varrho_4(\epsilon)$ are displayed in Table \ref{tab:uncertainty_exp_asym_coup_values}. Taking a linear regression of the log-log plot of $\varrho_2(\epsilon)$ \footnote{We neglect the first two points because we are interested in the limit $\epsilon\to 0$.}, we get
\begin{equation}
    \log_2\varrho_2(\epsilon) = 0.314\log_2\epsilon + 0.196
\end{equation}
with an $R^2$ value of 0.986. This indicates that $\mathfrak{u}_2\approx 0.314$, and since this is less than one, final state uncertainty does indeed exist in $S_2'$. By Eq. \eqref{eq:uncertainty-exp-frac-dim}, the fractal dimension $d_2$ of $\Sigma\cap S_2'$ is
\begin{equation}
        d_2 = n - \mathfrak{u}_2 \approx 2 - 0.314 = 1.686.
\end{equation}
Observing the visualization of $S_2'$ in Fig. \ref{fig:asym_basin_bw}, we can see that some black regions have boundaries that appear to be smooth, meaning that these particular subsets of the basin boundary have fractal dimensions close to 1. This accounts for the fractal dimension of $\Sigma\cap S_2'$ not being exceptionally close to 2: these boundaries that appear smooth make the overall basin boundary in this slice less ``rough.'' 

Now, considering the function $\varrho_4(\epsilon)$, we run a linear regression on the values $\log_2\varrho_4(\epsilon)$ vs. $\log_2\epsilon$ from Table \ref{tab:uncertainty_exp_asym_coup_values}, which yields
\begin{equation}
    \log_2\varrho_4(\epsilon) = 0.037\log_2\epsilon - 1.341
\end{equation}
with an $R^2$ value of 0.967. This indicates $\mathfrak{u}_4\approx 0.037$, which is exceedingly small and indicates extreme final state sensitivity. To put this in perspective, the fact that $\varrho_4(\epsilon)$ is proportional to $\epsilon^{\mathfrak{u}_4}$ means that to reduce the uncertainty regarding which attractor an initial state in $S_4'$ will be attracted to by a factor of 10, we will need to reduce the initial uncertainty $\epsilon$ by a factor on the order of $10^{27}$. By Eq. \eqref{eq:uncertainty-exp-frac-dim}, the fractal dimension $d_4$ of the basin boundary $\Sigma\cap S_4'$ is
\begin{equation}
    d_4 = n - \mathfrak{u}_4 \approx 4 - 0.037 = 3.963.
\end{equation}
This indicates that this basin boundary is extremely fractal; it has a comparable geometry to a true four-dimensional object even though it divides four-dimensional space. Even though Figs. \ref{fig:asym_basin_bw_y} and \ref{fig:3d-basin} don't show the entire four-dimensional set $S_4'$, they justify this high fractal dimension, where it is clear that the boundary dividing the seemingly random distribution of white and black points must be close to the dimension of the system's state space. 

We will conclude this section with a short discussion on how the uncertainty exponents $\mathfrak{u}_2$ and $\mathfrak{u}_4$ change as we expand out from $S_2'$ and $S_4'$. Although the basin classification method we used to classify the white and black basins of this asymmetrically coupled Rulkov neuron system doesn't take into account basin boundaries, we conjecture that $\mathfrak{u}_2$ approaches 1 as the bounds of $x_1$ and $x_2$ are expanded away from the set $S_2'$ because the white basin dominates far away the attractors. Additionally, we conjecture that $\mathfrak{u}_4$ stays relatively constant as the bounds are expanded from $S_4'$ because the white and black basins are both Class 2, so we suspect that they remain similarly overlapped with each other. These conjectures are supported by our computational experiments for different domains of $x$ and $y$ values.

\section{Conclusions}
\label{sec:conclusions}

The primary objective of this study was to analyze, classify, and quantify the geometries of the attractors, basins of attraction, and fractal basin boundaries of an asymmetrically electrically coupled system of two identical nonchaotic Rulkov neurons. We discovered the existence of a quasimultistability in the system, with a true nonchaotic spiking attractor and a chaotic spiking-bursting pseudo-attractor. Using box counting, we found that the chaotic pseudo-attractor was fractal but that its fractal dimension didn't coincide with its Lyapunov dimension. We also classified the basins of attraction of the nonchaotic spiking attractor and the chaotic pseudo-attractor. We found that in a two-dimensional slice of the basins near the slow variable values of the attractors, the basin of the chaotic pseudo-attractor took up almost all of the state space slice, while the basin of the nonchaotic spiking attractor had a finite measure and occupied increasingly small fractions of state space far away from its attractor. However, in all of four-dimensional state space, we discovered that both basins occupied fixed fractions of it. Finally, analyzing the system's uncertainty exponents and basin boundaries, we found extreme final state sensitivity requiring initial state uncertainty improvements on the order of $10^{27}$ to improve final state uncertainty by a factor of 10, as well as fractal basin boundaries with dimension $d\approx 3.96$ taking up almost all of four-dimensional state space.

We believe that our findings of complex and interesting geometries in the basins of attraction of this asymmetrically coupled Rulkov neuron system can be applied to other neuron models and dynamical systems, and we hope to see further study done in the classification of basins of attraction and quantification of fractal basin boundaries and final state sensitivity in the context of continuous-time and discrete-time neuronal systems. We also suggest research be done in using the method of basin entropy \cite{basin-entropy} to analyze the final state uncertainty and basin geometries of neuron systems, which extends this work. Additionally, we suggest further research be done into the existence of quasimultistability and pseudo-attractors in dynamical systems, which are concepts we found to yield rich and complex dynamics and geometries in our study. 

We conclude this paper with a discussion of the biological relevance of our results. Multistability is a prominent and important property in biological neuron systems that has been observed in a number of experimental studies \cite{lee, heyward, loewenstein}. One of the reasons why multistability is important for neuron function and living organisms is that its presence in cells can result in multipotentiality \cite{pisarchikbook}, which is an ability of certain cells to develop into multiple distinct cell types. This property can be modeled by a dynamical attraction, such as with cancer attractors \cite{huang, li-cancer}. The multistability we observed emerging from the interaction of neurons in this study has been more broadly found to be involved in the pathogenesis of neurological diseases such as Parkinson's disease \cite{tass, bergman}. As one of the first studies on the emergence of final state sensitivity and complex basin geometry in neuronal systems, the results detailed in this paper could have important biological significance, with minute variations in the state of biological neurons leading to vastly different functional outcomes through mechanisms of multipotentiality or pathogenesis. Therefore, building on existing experimental studies \cite{elson, abarbanel, varona}, we encourage exploration in experimentally replicating our results regarding final state sensitivity and complex basin geometry with coupled biological neurons.

\appendix

\section{Jacobian Matrix Partition}
\label{appx:partition}

Partitioning the $4\times 4$ Jacobian matrix $J(\mb{X})$ into four $2\times 2$ submatrices for all the possible piecewise parts of $\mb{F}(\mb{X})$ yields only five distinct forms, three of which appear on the diagonals and two of which appear on the off-diagonals. We will notate these $2\times 2$ matrices as follows:
\begin{align}
    J_{\text{dg},\,1}(x,\,\alpha,\,g;\,\mu) &= \begin{pmatrix}
        \alpha/(1-x)^2-g & 1 \\
        -\mu(1+g) & 1
    \end{pmatrix}, \\
    J_{\text{dg},\,2}(x,\,\alpha,\,g;\,\mu) &= \begin{pmatrix}
        -g & 1 \\
        -\mu(1+g) & 1
    \end{pmatrix}, \\
    J_{\text{dg},\,3}(x,\,\alpha,\,g;\,\mu) &= \begin{pmatrix}
        0 & 0 \\
        -\mu(1+g) & 1
    \end{pmatrix}, \\ 
    J_{\text{odg},\,1}(g;\,\mu) &= \begin{pmatrix}
        g & 0 \\
        \mu g & 0
    \end{pmatrix}, \\
    J_{\text{odg},\,2}(g;\,\mu) &= \begin{pmatrix}
        0 & 0 \\
        \mu g & 0
    \end{pmatrix}.
\end{align}
Now, we assign values to some parameters $a$, $b$, $c$, and $d$ based on which piecewise fast map interval $x_{1}$ and $x_{2}$ lie:
\begin{equation}
    \begin{cases}
        x_{1}\leq 0 & \implies a=1,\,b=1 \\
        0<x_{1}<\alpha_1+y_{1}+\mathfrak{C}_1 & \implies a=2,\,b=1 \\
        x_{1}\geq\alpha_1+y_{1}+\mathfrak{C}_1 & \implies a=3,\,b=2
    \end{cases},
    \label{eq:jacobian-abcd-1}
\end{equation}
\begin{equation}
    \begin{cases}
        x_{2}\leq 0 & \implies d=1,\,c=1 \\
        0<x_{2}<\alpha_2+y_{2}+\mathfrak{C}_2 & \implies d=2,\,c=1 \\
        x_{2}\geq\alpha_2+y_{2}+\mathfrak{C}_2 & \implies d=3,\,c=2
    \end{cases}.
    \label{eq:jacobian-abcd-2}
\end{equation}
Using this, we can compactify $J(\mb{X})$ into the form shown in Eq. \eqref{eq:jacobian}.

\bibliography{refs}

\end{document}